\documentclass{aa}  
\usepackage{courier,graphicx,epsfig,color,colordvi,natbib,url,twoopt,ulem,multirow}
\usepackage[varg]{txfonts}
\usepackage[breaklinks=true, draft]{hyperref} 
\usepackage{pdfcomment,acronym}

\bibpunct{(}{)}{;}{a}{}{,}    

\makeatletter
\newcommand{\bibnote}[2]{\@namedef{#1note}{#2}}
\newcommand{\biblink}[2]{\@namedef{#1link}{#2}}
\makeatother

\makeatletter
 \newcommandtwoopt{\citeads}[3][][]{%
   \nonstopmode
   \href{http://adsabs.harvard.edu/abs/#3}%
        {\def\hyper@linkstart##1##2{}%
         \let\hyper@linkend\@empty\citealp[#1][#2]{#3}}
   \biblink{#3}{\href{http://adsabs.harvard.edu/abs/#3}{ADS}}%
   \errorstopmode}            
 \newcommandtwoopt{\citepads}[3][][]{%
   \nonstopmode
   \href{http://adsabs.harvard.edu/abs/#3}%
        {\def\hyper@linkstart##1##2{}%
         \let\hyper@linkend\@empty\citep[#1][#2]{#3}}
   \biblink{#3}{\href{http://adsabs.harvard.edu/abs/#3}{ADS}}%
   \errorstopmode}            
 \newcommandtwoopt{\citetads}[3][][]{%
   \nonstopmode
   \href{http://adsabs.harvard.edu/abs/#3}%
        {\def\hyper@linkstart##1##2{}%
         \let\hyper@linkend\@empty\citet[#1][#2]{#3}}
   \biblink{#3}{\href{http://adsabs.harvard.edu/abs/#3}{ADS}}%
   \errorstopmode}            
 \newcommandtwoopt{\citeyearads}[3][][]{%
   \nonstopmode
   \href{http://adsabs.harvard.edu/abs/#3}%
        {\def\hyper@linkstart##1##2{}%
         \let\hyper@linkend\@empty\citeyear[#1][#2]{#3}}
   \biblink{#3}{\href{http://adsabs.harvard.edu/abs/#3}{ADS}}%
   \errorstopmode}            
\makeatother

\def\specchar#1{{\sc{#1}}}    

\def\vlos{\hbox{$v_{\rm{los}}$}}
\def\vturb{\hbox{$v_{\rm{micro}}$}}
\def\Blos{\hbox{$B_{\parallel}$}}

\def\Bhor{\hbox{$B_{\bot}$}}

\def\Bazi{\mbox{$\varphi$}}

\def\ltau{\hbox{log $\tau_{500}$}}
\def\dltau{\hbox{$\Delta$log $\tau_{500}$}}
\def\Br{\hbox{$B_{r}$}}

\def\Brtp{\mbox{$\vec{B}(r,\theta,\phi)$}}

\def\CRISP{CRisp Imaging SpectroPolarimeter}
\def\SST{Swedish 1-m Solar Telescope}

\def\Hinode{{\it Hinode\/}}
\def\AIA{{\it Atmospheric Imaging Assembly\/}}
\def\HMI{{\it Helioseismic and Magnetic Imager\/}}
\def\SDO{{\it Solar Dynamics Observatory\/}}

\def\stic{STockholm Inversion Code}
\def\ie{i.e.}
\def\eg{e.g.}

\def\CaIR{\mbox{Ca\,\specchar{ii}\,\,8542\,\AA}} 
\def\CaII{\mbox{Ca\,\specchar{ii}}}

\def\deg{\hbox{$^\circ$}}       

\def\FeI{\mbox{Fe\,\specchar{i}}} 
\def\FeIHMI{\mbox{Fe\,\specchar{i}\,\,6173\,\AA}} 
\def\Halpha{\mbox{H\hspace{0.1ex}$\alpha$}} 

\newcommand{\fov}[2]{{{#1}\arcsec$\times${#2}\arcsec}}
\newcommand{\XYis}[2]{\mbox{$(X,Y)=({#1}\arcsec,{#2}\arcsec)$}}

\newcommand{\dlambda}[1]{\mbox{$\Delta \lambda = {#1}$}}

\usepackage{xcolor}

\begin{document}

\title{Active region chromospheric magnetic fields}
\subtitle{Observational inference
versus magnetohydrostatic modelling}
\titlerunning{Active region chromospheric fields: inversions versus MHS modelling}
 
\author{G.~J.~M.~Vissers$^{1}$ \and 
S.~Danilovic$^{1}$ \and
X.~Zhu$^{2,3}$ \and
J.~Leenaarts$^{1}$ \and
C.~J.~D{\'i}az Baso$^{1}$ \and
J.~M.~da Silva Santos$^{1}$ \and
J.~de la Cruz Rodr{\'i}guez$^{1}$ \and
T.~Wiegelmann$^{3}$
}
\institute{Institute for Solar Physics, Department of Astronomy, 
 Stockholm University, AlbaNova University Centre,
 106 91 Stockholm, Sweden
 \and
 CAS Key Laboratory of Solar Activity, National Astronomical Observatories, Chinese Academy of Sciences, Beijing, China
 \and
 Max Planck Institute for Solar System Research,
 Justus-von-Liebig-Weg 3, 
 37077 G{\"o}ttingen, Germany
}

\date{}

\abstract
  {
  A proper estimate of the chromospheric magnetic fields is believed to improve
  modelling of both active region and coronal mass ejection evolution.
  However, since the chromospheric field is not regularly obtained for
  sufficiently large fields-of-view, estimates thereof are commonly obtained
  through data-driven models or field extrapolations,
  based on photospheric boundary conditions
  alone and involving pre-processing that may reduce details and dynamic range
  in the magnetograms.
  }
  {We investigate the similarity between the chromospheric magnetic field that is
 directly inferred from observations and the field obtained from a magnetohydrostatic
  (MHS) extrapolation based on a high-resolution photospheric magnetogram.
  }
   {Based on Swedish 1-m Solar Telescope \FeIHMI\ and
   \CaIR\ observations of NOAA active region 12723,
   we employed the spatially-regularised weak-field approximation (WFA) to
   derive the vector magnetic field in the chromosphere from \CaII, as well as
   non-LTE inversions of \FeI\ and \CaII\ to infer a model atmosphere for
   selected regions. Milne-Eddington inversions of \FeI\ serve as photospheric boundary
   conditions for the MHS model that delivers the three-dimensional field, gas
   pressure and density self-consistently.
   }
  {For the line-of-sight component, the MHS chromospheric field generally agrees
  with the non-LTE inversions and WFA, but tends to be weaker than those when larger in magnitude than 300\,G.
  The observationally inferred transverse component is systematically stronger,
  up to an order of magnitude in magnetically weaker regions, yet the
  qualitative distribution with height is similar to the MHS results.
  For either field component the MHS chromospheric field lacks the fine
  structure derived from the inversions.
  Furthermore, the MHS model does not recover the magnetic imprint
  from a set of high fibrils connecting the main polarities.
  }
  {The MHS extrapolation and WFA provide a qualitatively similar chromospheric
  field, where the azimuth of the former is better aligned with \CaIR\ fibrils
  than that of the WFA, especially outside strong-field concentrations.
  The amount of structure as well as the transverse field strengths are,
  however, underestimated by the MHS extrapolation.
  This underscores the importance of considering a chromospheric magnetic field
  constraint in data-driven modelling of active regions, particularly in the
  context of space weather predictions.
  }
\keywords{Sun: activity -- Sun: chromosphere -- Sun: photosphere -- Sun: magnetic
topology -- Radiative transfer}
\maketitle

\section{Introduction}\label{sec:introduction}
Our monitoring capabilities of solar activity drastically improved with the 24/7
coverage from the photosphere to the corona that the
\SDO\ (SDO;
\citeads{2012SoPh..275....3P}) 
and its two instruments, the 
\AIA\ (AIA;
\citeads{2012SoPh..275...17L}), 
and \HMI\ (HMI;
\citeads{2012SoPh..275..207S}, 
\citeads{2012SoPh..275..229S}) 
afford since early 2010.
SDO/HMI provides full-disk photospheric vector-magnetograms that have
been used as basis for field extrapolations and bottom boundary conditions
in data-driven modelling of, for instance, flare-productive active regions
(\eg\
\citeads{2016ApJ...820...16J},
\citeads{2020A&A...644A..28P}).
Similar observations are not currently obtained on a
routine basis for the chromospheric magnetic field\footnote{The SOLIS (Synoptic
Optical Long-term Investigations of the Sun; 
\citeads{2003SPIE.4853..194K})  
Vector Spectromagnetograph did provide near-daily full-disk line-of-sight
chromospheric magnetograms using \CaIR\ until late October 2017, but has not
been operational since.}, 
while several studies indicate that including the chromospheric field vector is
benificial both for data-driven modelling and recovering chromospheric and
coronal magnetic field structures in extrapolations (\eg\ 
\citeads{2009ApJ...696.1780D}, 
\citeads{2019ApJ...870..101F}, 
\citeads{2020ApJ...890..103T}). 
A chromospheric field constraint on erupting flux ropes could also aid in
modelling and forecasting coronal mass ejection (CME) evolution, thereby likely
improving the predictions for the geo-effectiveness of CMEs
\citepads{2019SpWea..17..498K}. 

In the absence of chromospheric constraints, data-driven modelling relies on
magnetic (and electric) field boundary conditions from the photosphere alone
(but see
\citeads{2008SoPh..247..269M}, 
\citeads{2008SoPh..247..249W}).
Depending on the modelling approach, the HMI magnetograms are passed through
several pre-processing steps that serve to \eg\ avoid flux-imbalance and nudge
the bottom boundary to a force-free state or ensure numerical stability,
but these may also impact the modelled field configuration in the higher
atmosphere.
For instance,
\citetads{2021A&A...645A...1V}, 
analysing an X2.2 flare, report discrepancies that can in part be attributed to
the pre-processing that smoothes and modifies the photospheric field, but they
also point out the importance of higher resolution in the modelling to uncover
details in the chromospheric atmospheric field configuration.
Both likely affect the photospheric and chromospheric field estimates and may
thus have implications for the space-weather modelling that builds on such
data-driven models.
The magnetohydrostatic (MHS) model (%
\citeads{2018ApJ...866..130Z}, 
\citeyearads{2019A&A...631A.162Z}, 
\citeads{2020A&A...640A.103Z}) 
that we consider here includes part of the aforementioned pre-processing, but uses the
original photospheric magnetogram as bottom boundary condition while iteratively
reaching a solution, which could yield a better estimate of the chromospheric
field configuration.

With current instrumentation the field-of-view (FOV) for
high-resolution chromospheric observations is often constrained to some
50\arcsec--70\arcsec\ on the side, meaning that larger active regions (or flux
ropes for that matter) cannot typically be contained entirely within a single
pointing, thereby limiting the benefits from incorporating such observations as
boundary conditions for modelling larger-scale structures.
A solution to this is to canvas extended regions by mosaicking
(\eg\
\citeads{2012AAS...22020111R}, 
\citeads{2013OptEn..52h1603H}), 
but such observations have not been common practice at the \SST\ (SST;
\citeads{2003SPIE.4853..341S}). 
Indeed, only one such study exists
\citepads{2014ApJ...797...36S}, 
presenting a 9$\times$6 mosaic that covers an active region with a FOV of
\fov{280}{180} in \Halpha\ imaging spectroscopy.
However, recent improvements in the SST pointing control software have now
rendered mosaicking a relatively simple task.

Here we analyse a flux-balanced mosaic of NOAA active region (AR) 12723 and
compare its chromospheric magnetic field as inferred from the observations with
that from a MHS model based on a high-resolution photospheric vector
magnetogram.
A spatially-regularised weak-field approximation (WFA) delivers the full-FOV
chromospheric magnetic field, while we use Milne-Eddington (ME) inversions to
obtain the photospheric field vector, which also serves as boundary condition
for the MHS extrapolation model. 
Furthermore, we perform non-LTE inversions of selected regions-of-interest
(ROIs) to allow a depth-stratified comparison with the MHS model.

The remainder of this publication is structured as follows.
Section~\ref{sec:observations} introduces the observations, mosaic construction
and post-processing to prepare the data for inversions.
Section~\ref{sec:bfield_inference} describes the ME, WFA and non-LTE inversion
techniques used to infer the photospheric and chromospheric magnetic field
vectors, as well as the MHS model setup.
We present our results in Section~\ref{sec:results}, considering the field
vector both for several regions of interest and on larger, active region scales.
Finally, we discuss our findings in
Section~\ref{sec:discussion} and present our conclusions in 
Section~\ref{sec:conclusions}.

\begin{figure*}[bht]
  \centerline{\includegraphics[width=\textwidth]{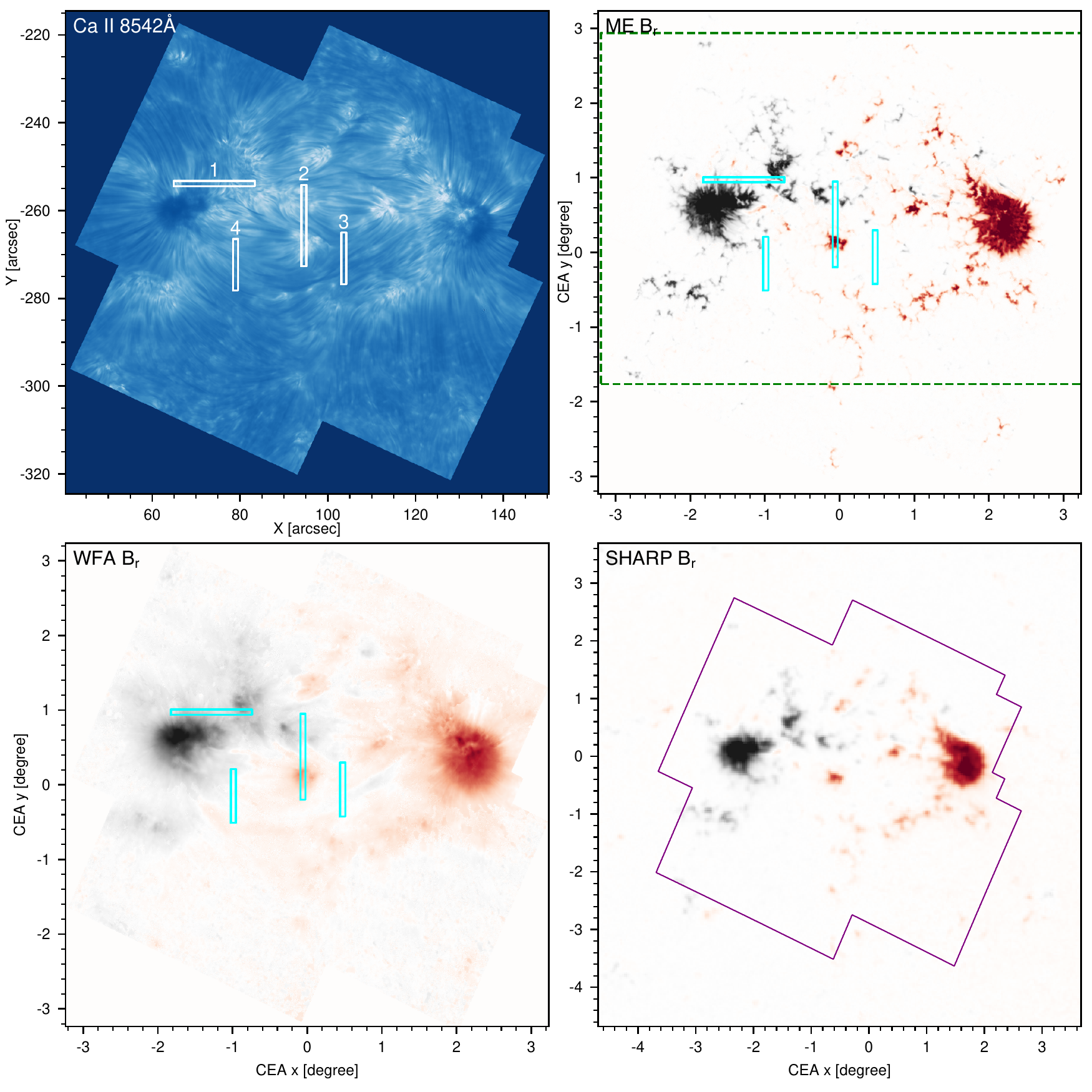}}
  \vspace{-2.5ex}
  \caption[]{\label{fig:fovs} %
    SST/CRISP mosaic and context SDO/HMI-derived SHARP magnetic field of NOAA AR
    12723.
    {\it Upper left\/}: \CaIR\ line core image.
    {\it Upper right\/}: Milne-Eddington inferred photospheric \Br\ 
    magnetic field from \FeI\,\,6173\,\AA\ observations by SST/CRISP. 
    {\it Lower left\/}: Chromospheric radial magnetic field \Br\ from the
    spatially-regularised weak-field approximation. 
    {\it Lower right\/}: Same as upper right, but from SDO/HMI.
    The line-of-sight magnetic field maps have been clipped to $\pm$1.5\,kG and
    are all reprojected and mapped to cylindrical equal-area (CEA) coordinates.
    The mapping to CEA precludes any direct comparison by coordinate of
    features in the top left panel versus the other panels.
    The white boxes in the top left panel indicate narrow regions-of-interest
    (ROIs) selected for (non-)LTE inversion and are indicated (approximately,
    \ie\ without mapping distortion) by cyan boxes in the top right and bottom
    left panels.
    The dashed green box in the top right panel highlights the FOV selected for
    the magnetohydrostatic simulation, while the purple contour in the lower
    right panel traces the SST mosaic field-of-view.
  }
\end{figure*}

\section{Observations and reduction}\label{sec:observations}
On 30 September 2018 we observed NOAA AR 12723 with the
the \CRISP\ (CRISP; 
\citeads{2008ApJ...689L..69S}) 
and CHROMIS 
\citepads{2017psio.confE..85S} 
at the SST. 
These observations consist of two rounds of four overlapping pointings in a
2$\times$2 pattern for a context mosaic, followed by a regular, single-pointing
time series.
For this study, we focus only on the context data and selected the set of
pointings with the highest image contrast to construct the mosaic that we
analyse here.

The mosaic has a mean observing time of 09:24:03\,UT, is centred at
\XYis{95}{-270}, corresponding to $\mu=0.95$, and covers about \fov{110}{110} at 0\farcs{058}\,pix$^{-1}$.
Figure~\ref{fig:fovs} presents the mosaic in \CaII\ (in helioprojective
representation), with photospheric and
chromospheric vertical magnetic field maps and context magnetic field from an
SDO HMI Active Region Patch (SHARP;
\citeads{2014SoPh..289.3549B}) 
in cylindrical equal-area (CEA) projection.

Imaging spectropolarimetry was obtained in the \FeI\,\,6173\,\AA\ and \CaIR\
lines.
\FeI\ was sampled at eight different wavelengths, equidistantly
distributed at 40\,m\AA\ spacing between $-$0.12\,\AA\ and +0.16\,\AA. 
The \CaII\ sampling covers $\pm$0.95\,\AA, with finer spacing in the core (down
to 70\,m\AA) and coarser spacing in the outer wings (up to 350\,m\AA).

We used the SSTRED pipeline 
\citepads{2015A&A...573A..40D,2018arXiv180403030L} 
to reduce the data, which includes image restoration using Multi-Object
Multi-Frame Blind Deconvolution (MOMFBD;
\citeads{2005SoPh..228..191V}). 
Remaining small-scale seeing-induced deformations
were removed according to the recipe by
\citetads{2012A&A...548A.114H}, 
while further destretching to correct for rubber-sheet seeing effects was
performed following
\citetads{1994ApJ...430..413S}. 
We minimised the fringes in Stokes $Q$, $U$ and $V$ by Fourier-filtering
(following the procedure in 
\citeads{2020A&A...644A..43P}) 
and increased the signal-to-noise (S/N) in the \CaIR\ Stokes $Q$ and $U$ by
applying the denoising neural network of
\citetads{2019A&A...629A..99D}. 

The mosaic was then constructed by aligning these cleaned and denoised
pointings at sub-pixel accuracy through cross-correlation. 
To avoid sharp edge-effects from the individual pointings, we applied so-called
feathering in the overlapping regions, combining the pointings according to
a linear weight map that ensures a smooth transition between them.
In addition, we smoothed the data across the pointing edges over a width of
10\,pixels (\ie\ about 0\farcs{6}) to further reduce any remaining edge-effects.
These mosaic construction steps were determined based on the \FeI\ and \CaII\
wide-band images and then applied to the respective spectropolarimetric data.
Finally, we cut away two small patches at about \XYis{145}{-245} and
\XYis{145}{-275} as both contained a strong artefact that would otherwise have led
to spurious results in the inversions and as such in the lower boundary
conditions for the MHS model.


%
CRISPEX\footnote{https://github.com/grviss/crispex} (%
\citeads{2012ApJ...750...22V}, 
\citeads{2018arXiv180403030L}) 
was used to verify the mosaic alignment and for data browsing.

\section{Magnetic field inference}\label{sec:bfield_inference}
\subsection{Approximations for field-of-view maps}\label{sec:bfield_mewfa}
With over 2.3 million data pixels in the mosaic, performing a full-FOV non-LTE
inversion would be computationally prohibitive.
In order to nonetheless obtain a field-of-view estimate of the magnetic field
vector we performed Milne-Eddington (ME) inversions of the \FeI\ data and
applied a
spatially-regularised weak-field approximation (WFA) to the \CaII\ data.

ME inversions are the workhorse of pipelines routinely delivering
photospheric magnetograms for, \eg, SDO/HMI and \Hinode\ SOT/SP, and
(in the past) SOLIS/VSM.
For our ME inversions of the SST/CRISP \FeI\,\,6173\,\AA\ data we used 
{\tt pyMilne}\footnote{https://github.com/jaimedelacruz/pyMilne}, a hybrid implementation written
in Python and C++ 
\citepads[][see the upper-right panel of Fig.~\ref{fig:fovs}]{2019A&A...631A.153D}.

The WFA, on the other hand, is often used to obtain an estimate of the
chromospheric magnetic field vector (\eg\ 
\citeads{2013A&A...556A.115D}, 
\citeads{2007ApJ...663.1386P}, 
\citeads{2012SoPh..280...69H}, 
\citeads{2017ApJ...834...26K}). 
Here we use an extension of this method by 
\citetads{2020A&A...642A.210M}, 
{\tt spatial\_WFA}\footnote{https://github.com/morosinroberta/spatial\_WFA},
that imposes spatial continuity in the solution and that has recently been
applied with success to \CaIR\ observations of an X2.2 flare
\citepads{2021A&A...645A...1V}. 
The benefit of this spatially-regularised WFA is that, with well-chosen
parameters, the adverse effects of noise can be greatly reduced.

\subsection{Non-LTE inversions}
\label{sec:bfield_nlte}
We also perform (non-)LTE inversions of selected regions of interest (ROIs) of our \FeI\ and \CaII\
mosaic data using the \stic\ (STiC;
\citeads{2016ApJ...830L..30D}, 
\citeads{2019A&A...623A..74D}), 
which after settling on a best fit profile delivers an atmospheric model with
(among others) temperature and velocity stratification, as well as the magnetic
field vector.
STiC is an MPI-parallel non-LTE inversion code built around a modified version
of RH 
\citepads{2001ApJ...557..389U} 
to solve the atom population densities assuming statistical equilibrium and
plane-parallel geometry, using an equation of state extracted from the SME code 
\citepads{2017A&A...597A..16P}. 
We assumed complete frequency redistribution (CRD) in our inversions.
The radiative transport equation is solved using cubic Bezier solvers 
\citepads{2013ApJ...764...33D}. 
The \CaII\ inversions were performed in non-LTE using a 6-level calcium model
atom and assuming CRD, while \FeI\ was treated in LTE.

We assumed FAL-C as initial model atmosphere, spanning between $\ltau=0.1$ and
$-$7.1 at $\dltau=0.2$ spacing. 
For the magnetic field components \Blos, \Bhor\ and the azimuth \Bazi\ we used the
WFA results as initial guess.
The inversions were run in three cycles, with an increased number of nodes in
temperature, velocity and magnetic field for each subsequent cycle (see
Table~\ref{tab:cycles}).
Contrary to standard procedure, we inferred all three field components with two nodes
already in the first cycle, as the S/N difference between \FeI\ and \CaII\ in
Stokes $Q$, $U$ and $V$ otherwise tended to push the model atmosphere to a
photospheric field configuration that dominated the final results.
Spatial and depth smoothing between the cycles minimised the persistence of
poorly fitted pixels.
We present our inversion results in Section~\ref{sec:rois}.

\begin{table}
  \caption{Number of nodes used in each inversion cycle.}
\begin{center}
  
\begin{tabular}{l|ccccc}%
  \hline\hline
  {} & \multicolumn{5}{c}{Parameter} \\ \cline{2-6}
  Cycle & $T$ & \vlos & \vturb & \Blos & \Bhor\ and $\varphi$ \\
  \hline
  1 & 4 & 2 & 0 & $-$5.0, $-$0.5 & $-$5.0, $-$0.5 \\
  2 & 6 & 3 & 2 & $-$5.0, $-$0.5 & $-$5.0, $-$0.5 \\
  3 & 9 & 4 & 3 & $-$5.0, $-$2.0, $-$0.5 & $-$5.0, $-$0.5 \\
  \hline
\end{tabular}
\end{center}
\label{tab:cycles}
\end{table}

\subsection{Azimuth disambigution}\label{sec:disambig}
The magnetic field azimuth \Bazi\ recovered from the inference methods described
in Sections~\ref{sec:bfield_mewfa} and \ref{sec:bfield_nlte} contains a
180\deg-ambiguity that needs to be resolved to determine $\vec{B}(x,y,z)$.
Since we use the ME-inferred field as photospheric boundary condition for a
magnetohydrostatic model, 
we resolve this ambiguity in the ME-derived photospheric azimuth according to
the minimum energy method (MEM)
\citetads{1994SoPh..155..235M},  
using the implementation by 
\citetads{2014ascl.soft04007L}.  
Regions unstable for disambiguation were identified by running twenty instances
of the MEM code with different random number seeds and setting the azimuth of
those pixels for which the result changed by more than 45\deg\ by majority vote.

%

We did, however, not attempt to disambiguate the chromospheric WFA azimuth with
the same procedure as this re-introduced pixelisation of the FOV, obviating
benefits from spatial regularisation.
Also, given the small extent of the selected ROIs for non-LTE inversion, we did
not disambiguate the inferred azimuth.
In both cases we focus on the orientation, rather than the
direction of the transverse field in the $xy$-plane.

\subsection{Magnetic field decomposition and reprojection}
\label{sec:bfield_cea}
A challenge when comparing observations with numerical experiments is that the
observed quantities are projected with respect to the line-of-sight, where
pixels do not all cover the same physical extent on the Sun.
A common practice when using observations as boundary conditions for simulations
is therefore to reproject the observed magnetic field to cylindrical equal-area
(CEA) coordinates, \ie\ as if the target were observed top-down and with
equally-sized pixels. 
Indeed, one of the formats that SHARPs can be acquired in has the magnetic field
reprojected to CEA.

SolarSoft IDL\footnote{http://www.lmsal.com/solarsoft}
provides a pipeline that delivers the HMI data in CEA coordinate
projection and we have adapted this pipeline to handle SST data products. 
Our version is written in Python and can be found online\footnote{See the {\tt
remap} utility package in https://github.com/ISP-SST/ISPy}.
The resulting \Brtp\ maps have a 0.003\deg\ pixel-size (equivalent to about
0\farcs{05} at disc centre),
\ie\
ten times smaller than for the SHARP maps, given that the pixel size of the
SST observations is about a factor 10 smaller than that of HMI.
Samples of the photospheric and chromospheric $B_{r}$ maps are shown in
Fig.~\ref{fig:fovs}.

Finally, we also apply the inverse reprojection operation, \ie\ from CEA back
to the frame that the observations were obtained in, to facilitate comparison between
the MHS model and non-LTE inversion results (given that the latter were
performed in the original observations frame). 

%
%
%
%

\subsection{Magnetohydrostatic model}
\label{sec:mhs}
\subsubsection{Model setup and boundary conditions}
We use a cut-out of the ME-derived magnetic field vector in CEA coordinates (see
green dashed box in Fig.~\ref{fig:fovs}) as lower boundary condition for a
magnetohydrostatic (MHS) model of the active region, where the pixels outside
the SST mosaic FOV are filled with values from the SHARP.
The computational domain has 2144$\times$1568$\times$160 pixels of 36.5\,km on
the side, thereby providing the field over a surface of about 78$\times$57\,Mm,
for more than 5.8\,Mm in height.
This cut-out is well-balanced in flux, with a ratio of net-to-unsigned flux of
less than $2\times10^{-3}$.

The MHS model (%
\citeads{2018ApJ...866..130Z}, 
\citeyearads{2019A&A...631A.162Z}, 
\citeads{2020A&A...640A.103Z}) 
takes into account the interaction between the magnetic field and plasma, and is
appropriate for the lower atmosphere where the plasma $\beta$ (\ie\ ratio of gas
pressure to magnetic pressure) can be large.
The model is realised in two steps.
First, we initialise the magnetic field by performing a non-linear force-free
field (NLFFF) extrapolation following
\citetads{2004SoPh..219...87W}, 
based on a pre-processed version of the ME-derived magnetogram
\citepads{2006SoPh..233..215W}. 
The plasma is initially distributed along the magnetic field lines assuming
gravitational stratification.
Next, the code simultaneously computes the magnetic field vector, plasma
pressure and density, while iteratively minimising an objective function
(Eq.~(4) in \citeads{2018ApJ...866..130Z}). 
In this step, the original magnetogram is used as the bottom
boundary condition.

\begin{figure*}[ht]
  \includegraphics[width=\textwidth]{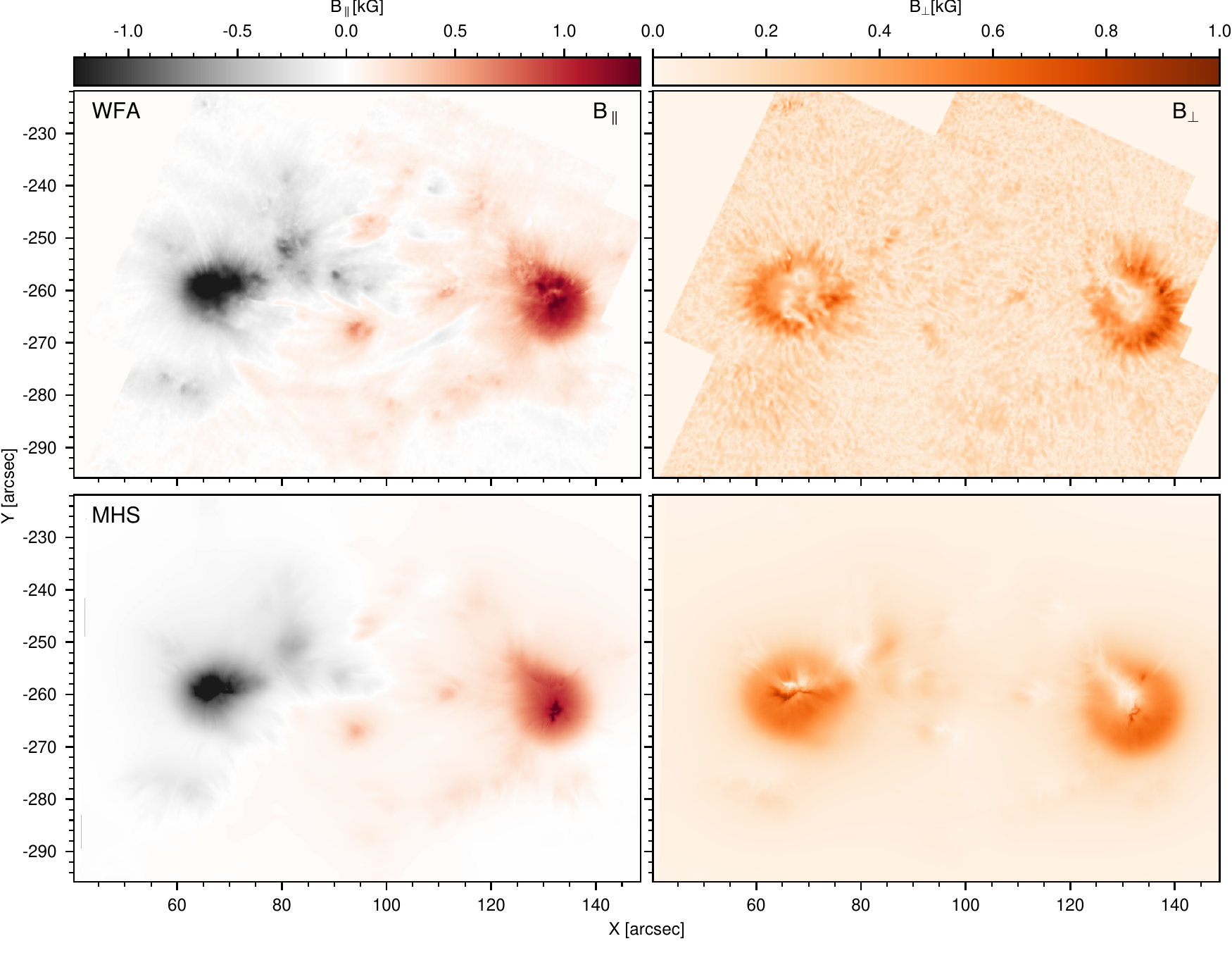}
  \caption[]{\label{fig:wfa_mhs} %
  Magnetic field components \Blos\ ({\it left\/}) and \Bhor\ ({\it right\/})
  from the WFA ({\it top row\/}) and from the MHS model ({\it
  bottom row\/}), averaged over the equivalent height range (as determined from
  the column mass) that the WFA is expected to be sensitive to.
  To allow direct comparison between the WFA and MHS results, the panels in each
  column have been clipped to the same values:  between $[-1.35,1.25]$\,kG for
  \Blos\ and $[0,1]$\,kG for \Bhor.
  In addition, the \Blos\ panels (left column) have been gamma-adjusted to
  improve visibility of weaker features.
  }
\end{figure*}

\subsubsection{Post-processing for comparison with inversions}
\label{sec:mhs_postproc}
Comparing the MHS model results to the non-LTE inversions of selected ROIs (or WFA of
the full field) requires post-processing, since the MHS field maps are in
CEA coordinates and geometrical $z$-height scale, rather than in helioprojective Cartesian 
$(X,Y)$ and optical depth \ltau.
Firstly, we therefore reproject the MHS magnetic field back to the same
frame as the observations, using the inverse operation. 
%
Secondly, we extract the chromospheric magnetic field at equivalent
heights in the MHS model and the non-LTE-inferred atmospheres.
To that end, we inspect the response functions of \CaIR\ Stokes $Q$, $U$ and $V$
for all ROIs, finding that the highest sensitivity in $Q$ and $U$ to changes in
\Bhor\ and in $V$ to changes in \Blos\ is reached between $\ltau=-4.1$ and
$-$5.1.
For our top-down views of the chromospheric field we therefore average the
inferred magnetic field components over this range and determine the column
mass equivalent to those \ltau-depths, which in turn yields the column mass
depths to average over in the MHS model.
In addition, we recompute the zero point of the column-by-column hydrostatic
$z$-scale in the inferred atmospheres by imposing horizontal gas pressure
equilibrium (including magnetic pressure) in both the $x$- and $y$-direction at
the continuum $\tau_{500}=1$ depth, shifting the $z$-scale of each column in the
ROI accordingly.
This procedure follows the one outlined in Section 4.3 of
\citetads{2020A&A...642A.210M}. 
As pointed out in that study,
this is necessarily only an approximation as we cannot
include the Lorentz force or magnetic pressure in our inversions.
Thirdly, and finally, as the MHS azimuth covers the full 360\deg\ circle, while
our WFA and inversion results retain a 180\deg-ambiguity, we introduce this
ambiguity in the MHS result by subtracting 180\deg\ from any azimuth equal to or
greater than 180\deg, so that the azimuth maps can be compared side-by-side.

\begin{figure*}
  \includegraphics[width=\textwidth]{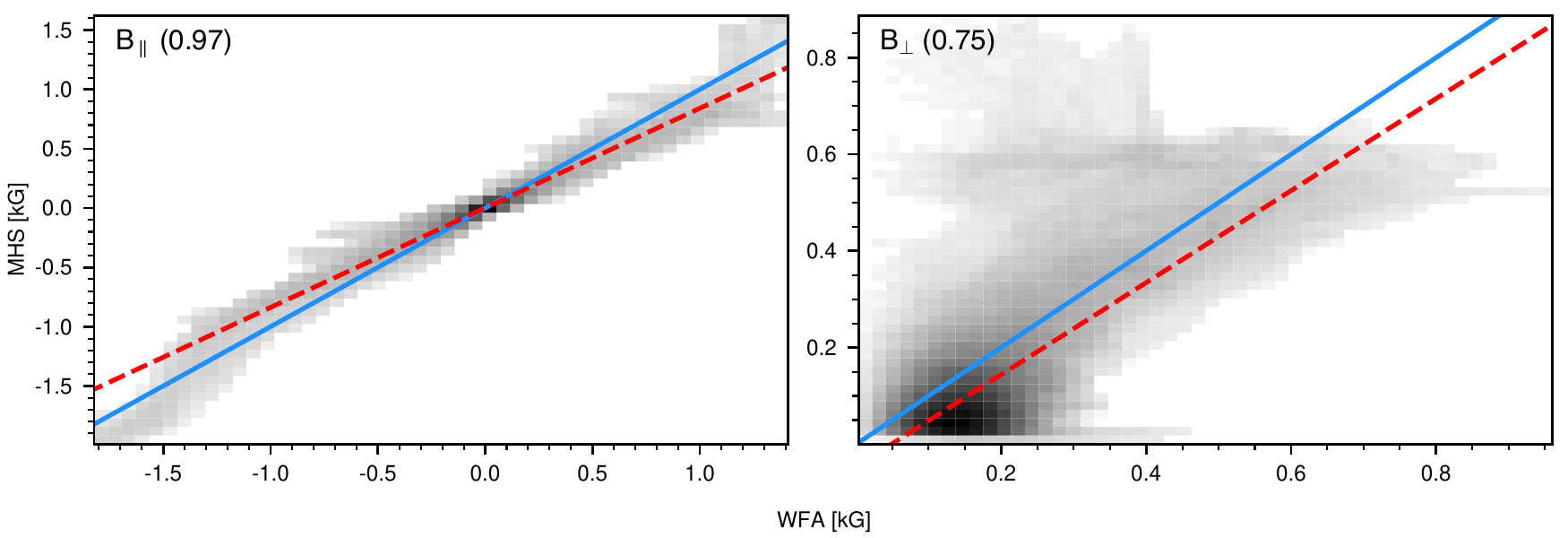}
  \caption[]{\label{fig:wfa_mhs_scatter} %
  Two-dimensional histograms of the WFA versus MHS magnetic field for \Blos\
  ({\it left\/}) and \Bhor\ ({\it right\/}) for the overlapping part of the FOV
  shown in Fig.~\ref{fig:wfa_mhs} (\ie\ excluding pixels without \CaII\ data).
  The diagonal blue line indicates a 1:1 relationship, while the dashed red line
  shows a linear fit to the data. 
  }
\end{figure*}

\section{The chromospheric magnetic field in AR 12723}
\label{sec:results}
An important question is whether the MHS model can reproduce the
chromospheric field, solely based on the photospheric boundary condition
provided by the ME inversion of our \FeI\ data.
To evaluate the degree of similarity, we compare the MHS results both on active
region scales with the spatially-regularised WFA field
(Sections~\ref{sec:wfa_mhs} and \ref{sec:fib_azim}) and in closer detail for
selected regions-of-interest that we inverted in non-LTE
(Section~\ref{sec:rois}).
While we inverted four ROIs, we focus on the two most interesting ones, ROIs 1
and 3  in the main part of this publication and include discussion of ROIs 2 and
4 in Appendix~\ref{sec:other_rois}.

\subsection{Magnitude of the magnetic field components}
\label{sec:wfa_mhs}
The WFA provides the three magnetic field components over a corrugated
surface in $z$-scale that depends on the location within the field-of-view,
which we need to take into account when comparing with the MHS field.
As described in Section~\ref{sec:mhs_postproc}, we determine the column mass
equivalent to $\ltau=-4.1$ and $-$5.1 and use those boundaries to select the
column mass depths over which to average in the MHS model.
Figure~\ref{fig:wfa_mhs} compares the \Blos\ and \Bhor\ maps from the
spatially-regularised WFA and the resulting depth-averaged MHS model. 

Qualitatively, there is good agreement for the strongest field concentrations,
both in \Blos\ and \Bhor, however much of the substructure that the WFA recovers
in \eg\ the superpenumbra of both sunspots, appears more as a homogeneous and
comparatively weaker-field haze in the MHS panels.
For instance, the strong \Blos\ inside the sunspots is more extended in the WFA
results, especially for the positive-polarity sunspot, and elongated
strong-field concentrations are seen to extend radially from both sunspots in
the WFA \Bhor\ that are entirely absent in the MHS.

Meanwhile, the stronger field concentrations outside the sunspots in the WFA
\Blos\ are only recovered in location in the MHS \Blos, but weaker in strength
and hardly in terms of fine-structure.
The lack of magnetic features outside the sunspots in the MHS model is
particularly striking when comparing both \Bhor\ panels (right-hand column),
although much of the fine-structure in the WFA \Bhor\ in the lower right of the
FOV (\ie\ left of $X=80$\arcsec\ and below $Y=-280$\arcsec) is likely spurious
as suggested by weak total linear polarisation in those areas.
As such, the MHS \Bhor\ field does a good job in identifying locations with
significant horizontal chromospheric field.
Another feature of interest is the band of negative \Blos\ around
\XYis{110}{-270} in the WFA panel, that coincides with dark elongated fibrils as
seen in the \CaII\ line core (cf.~\eg\ Fig.~\ref{fig:fovs} top left panel).
While it is only visible weakly because of its much smaller field strength
compared to the sunspots, it is evidently absent in the MHS panel.
We consider these high fibrils in further detail in Section~\ref{sec:roi3}.

Figure~\ref{fig:wfa_mhs_scatter} compares the WFA and MHS magnetic field
components quantitatively in two-dimensional histograms.
For \Blos\ (left-hand panel) the correlation is generally tight, as also
evidenced by the Pearson correlation number of 0.97, and indeed for weak field
strengths (below 0.2--0.3\,kG) this correlation appears to be essentially
one-to-one.
For larger field strengths the scatter cloud widens and overall tends to lower
field strengths in the MHS compared to the WFA, as indicated by the linear
regression (dashed red line).
Considering \Bhor (right-hand panel) the scatter cloud is much wider and while
the WFA-recovered field strengths are generally larger than in the MHS model
(\ie\ the darkest parts of the two-dimensional histogram are found right of the
one-to-one line), there is an extended cloud of points at WFA field strengths
below 0.5\,kG that exceed that value by some 0.35--0.70\,kG in the MHS model.
These points correspond to the spuriously looking narrow field enhancements in
both sunpsot centres in the MHS \Bhor\ map.
Careful examination of the data suggests that these regions appear as a result
of data processing and become pronounced only after averaging in height.
So, despite the wider scatter and offset by about +100\,G in favour of the WFA, the
linear regression has a slope of 0.96, indicating an essentially 1:1
relationship between the WFA and MHS in \Bhor.

\subsection{Chromospheric field azimuth and fibril orientation}
\label{sec:fib_azim}
\begin{figure*}[bht]
  \centerline{\includegraphics[width=\textwidth]{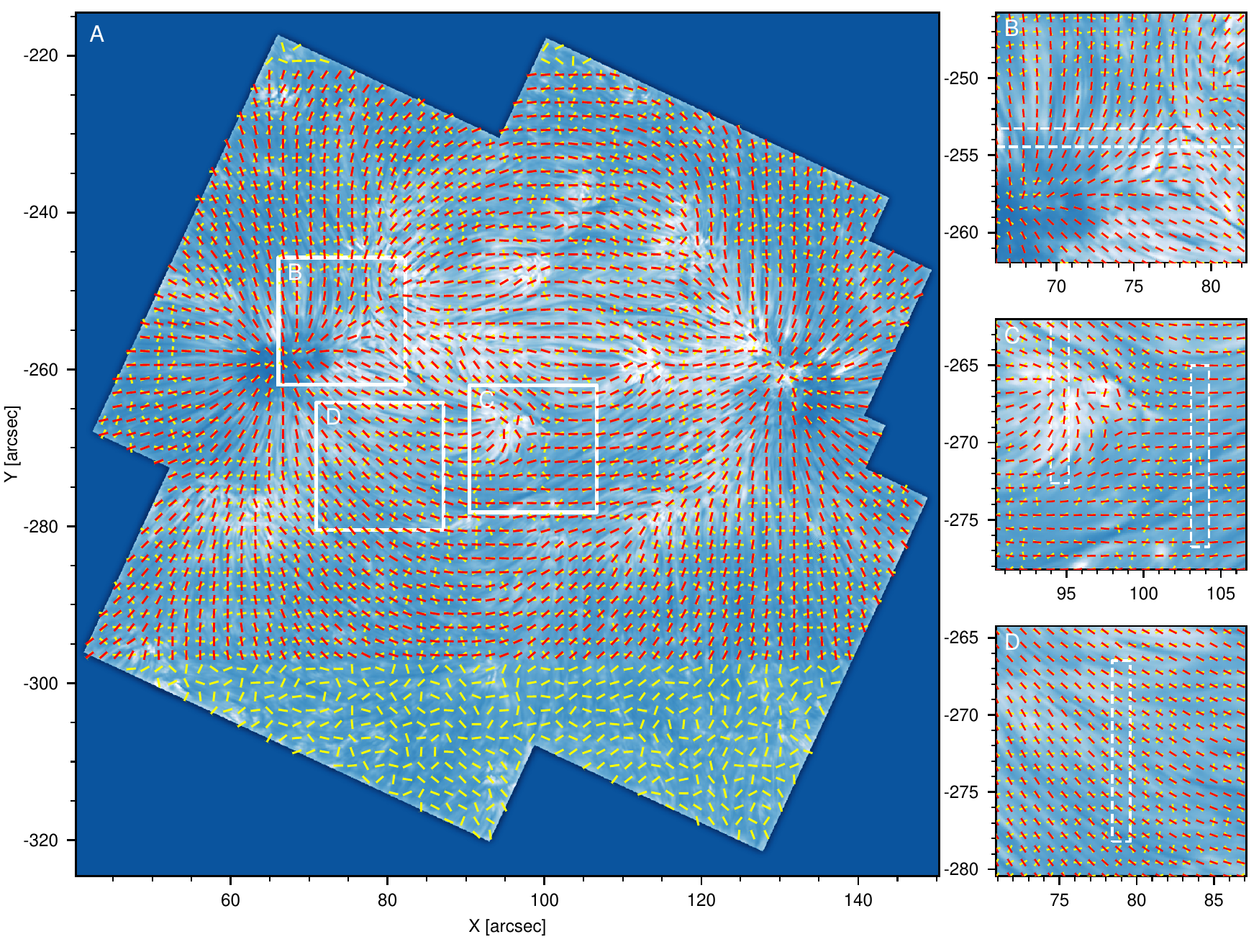}}
  \vspace{-2.5ex}
  \caption[]{\label{fig:fib_azim} %
  Comparison of fibril orientation and chromospheric field azimuth from the
  spatially-regularised WFA and the MHS model.
  The background \CaIR\ image has been unsharp masked with a radius of
  0\farcs{5} to enhance fibrillar structures.
  The azimuth is indicated by yellow (WFA) and red (MHS) bars, where arrowheads
  have been omitted since we did not resolve the 180\deg-ambiguity for the WFA.
  The three white boxes in the left-hand panel (a) indicate the selections
  presented as panels (b)--(d) in the right-hand column and have been chosen
  such as to contain part of the ROIs in Fig.~\ref{fig:fovs}, where panel (b)
  covers ROI 1, panel (c) ROIs 2 and 3, and panel (d) ROI 4.
  }
\end{figure*}

As the chromosphere is a low-$\beta$ regime where the magnetic pressure dominates
in the pressure balance, the fibrils visible in \eg\ the \CaIR\ line core are
expected to be relatively well-aligned with the local magnetic field vector,
even though the extent to which this is the case may vary depending on target
and local conditions (\eg\ 
\citeads{2011A&A...527L...8D}, 
\citeads{2013ApJ...768..111S}, 
\citeyearads{2015SoPh..290.1607S}, 
\citeads{2015ApJ...802..136L}, 
\citeads{2016ApJ...826...51Z}, 
\citeads{2016ApJ...831L...1M}, 
\citeads{2017A&A...599A.133A}, 
\citeads{2017ApJS..229...11J}, 
\citeads{2019A&A...631A..33B}). 
Figure~\ref{fig:fib_azim} shows the azimuth \Bazi\ derived from the WFA in
yellow and from the MHS model in red overlaid on an unsharp masked (with radius
0\farcs{5}) \CaII\ line core intensity image of AR 12723.
The MHS azimuth is obtained by averaging over the column mass range equivalent
to $\ltau=-4.1$ to $-$5.1 (see Section~\ref{sec:mhs_postproc}).
Given that we did not disambiguate the WFA azimuth (see also
Section~\ref{sec:disambig}) and since we are primarily interested in comparing
the fibril orientation with the field azimuth, we omit all arrowheads in the
Fig.~\ref{fig:fib_azim} overlays.

Considering the full FOV (panel A), the WFA azimuth agrees well with both the
MHS azimuth and the orientation of fibrillar structures in the superpenumbra of
both sunspots, as well as the fibrils in parts of the inter-spot region (\eg\
above and to the right of box C).
Moving away from the sunspot centres and the stronger field-concentrations and
pores, the WFA azimuth disagrees more often with the fibril orientation, \eg\
in part of the superpenumbra of the
negative-polarity sunspot (above and to the very left of box B), the fibrils
connected to bright plage south of that spot (to the lower left of box D) and
notably for the dark fibril around \XYis{128}{-245}, to the north of the
positive-polarity sunspot.
In contrast, the MHS azimuth is relatively well-aligned with the fibrils also at
greater distance from the sunspots (especially around the left-hand
negative-polarity sunspot and aforementioned dark fibril north of the
positive-polarity sunspot), which in turn means 
the WFA and MHS azimuth more often cross at near-perpendicular angles.
At the same time, the orientation of a long fibril that extends from the
positive-polarity sunspot into box C appears not to be captured in the MHS model
(the azimuth bars are essentially parallel to the $x$-axis) while the WFA
azimuth is better aligned with this fibril at several places.
Finally, outside the strong-field areas of the active region (\eg\ in the
bottom, lower right and upper left parts of the mosaic FOV) the WFA azimuth is
more random, but the lack of clear fibrillar structures also renders a
comparison in these regions less straightforward. A source of this randomness is that the WFA will always provide an azimuth as output, even if the linear polarization signal is completely dominated by noise.

Figure~\ref{fig:fib_azim}b--d zoom in on three cutout boxes of about
\fov{16}{16} marked B--D in Fig.~\ref{fig:fib_azim}a.
Cutout B covers part of the negative-polarity sunspot and its superpenumbra, as
well as the ROI 1 box (indicated by dashed white lines), cutout C a magnetic
concentration in the inter-spot region (including ROIs 2 and 3) and cutout D the
superpenumbra south of the negative-polarity sunspot (including ROI 4).
These detailed views paint a similar picture as the FOV overview, in that the
WFA azimuth is often well-aligned with \CaII\ fibrils at or in the vicinity of
strong field (\eg\ dark fibrils close to the sunspot in cutout B, fibrils
extending to the south from the positive-polarity pore around \XYis{94}{-270}
and those that cross the lower right quadrant of the box (between about
\XYis{93}{-278} and \XYis{106}{-269}) in cutout C, upper left part of cutout D
(between \XYis{80}{-270} and the upper left corner), but the mis-alignment
increases when moving away from sunspot or pore (\eg\ top 3\arcsec\ of cutout B,
fibrils north of the pore in cutout C, most of cutout D except for the upper
left quadrant).
In contrast, the MHS azimuth more often traces the fibrils over larger parts of
these sub-FOVs, with the notable and aforementioned exception of the long dark
fibrils crossing the lower right quadrant of cutout C.

\begin{figure*}[bht]
  \centerline{\includegraphics[width=\textwidth]{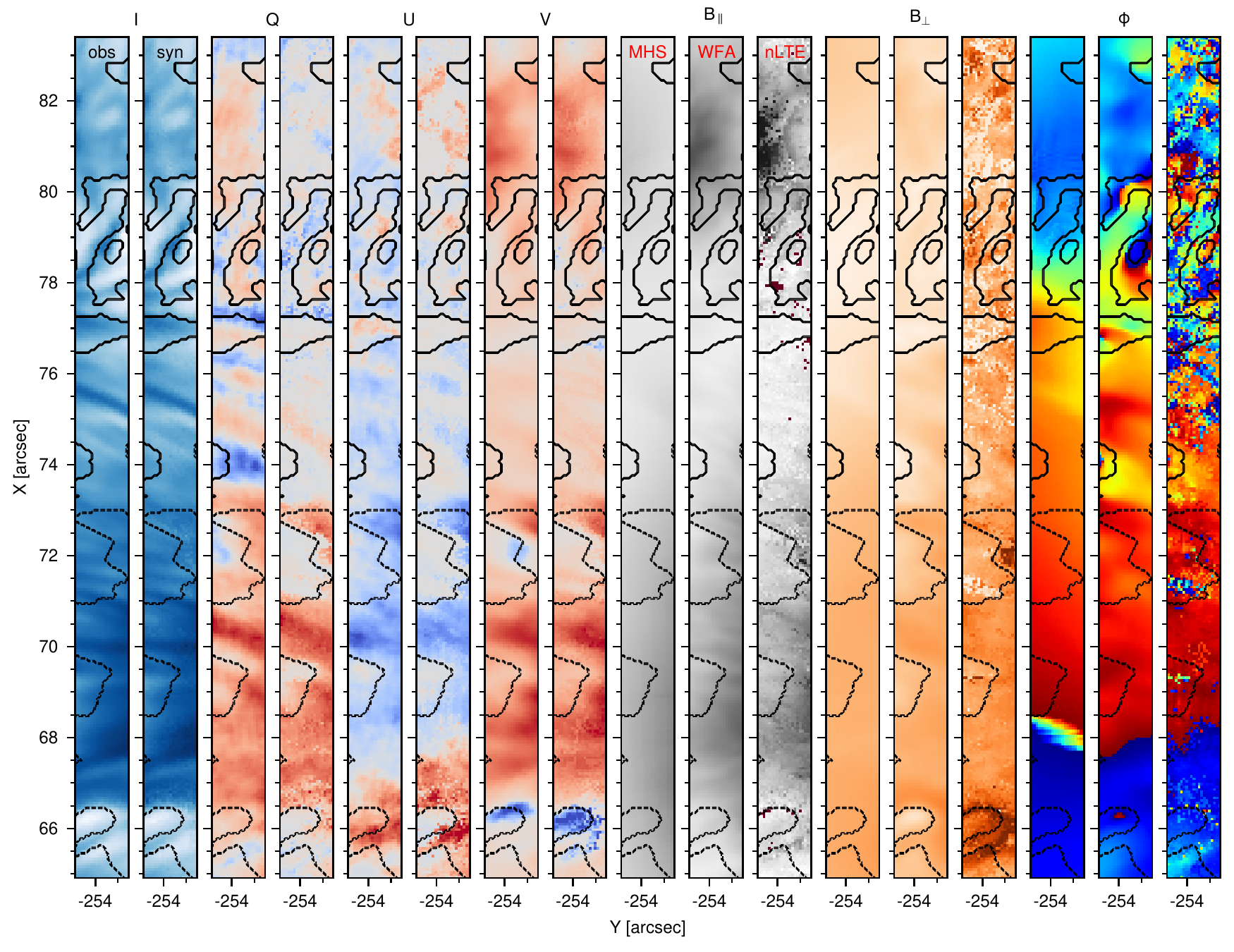}}
  \vspace{-2.5ex}
  \caption[]{\label{fig:roi1} %
  Maps of Stokes intensity and magnetic field components for ROI 1.
  The ROI has been rotated by 90\deg\ counter-clockwise to allow for an
  identical layout as Figs.~\ref{fig:roi3}, \ref{fig:roi2} and \ref{fig:roi4}.
  The first eight columns contain the Stokes $I$, $Q$, $U$ and $V$ intensity
  maps, where each column pair shows first the observed and then the synthetic
  maps, clipped to the same values.
  The Stokes $I$ map is taken at \CaIR\ line centre, while the other Stokes
  components are shown at \dlambda{+0.21\,\AA} offset.
  The remaining nine columns contain the magnetic field components \Blos, \Bhor\
  and azimuth $\varphi$, where each column triple shows from left to right the
  MHS, WFA and non-LTE inversion result, again clipped to the same values.
  The non-LTE magnetic field components are the average over
  $\ltau=[-5.1,-4.1]$, while the MHS field components are averaged over the
  equivalent $z$-height (determined per pixel).
  The contours highlight regions where the WFA-derived \Bhor\ transitions 150\,G
  ({\it solid\/}) and 350\,G ({\it dashed\/}).
  The tick spacing is the same for both the $x$- and $y$-axes. 
  }
\end{figure*}

\begin{figure*}[bht]
  \centerline{\includegraphics[width=\textwidth]{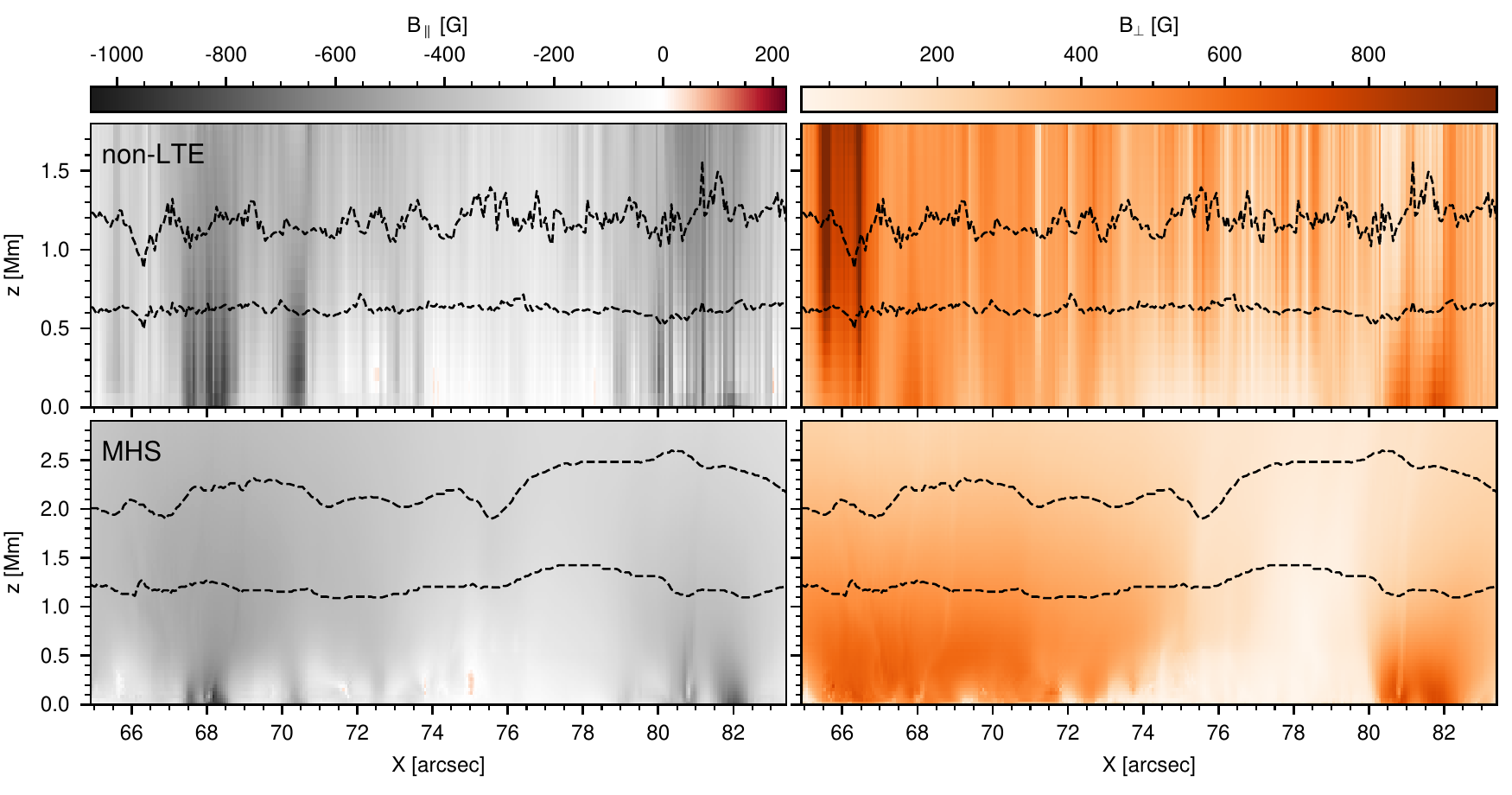}}
  \vspace{-2.5ex}
  \caption[]{\label{fig:vcut_roi1} %
  Vertical cuts in \Blos\ ({\it left column\/}) and \Bhor\ ({\it right
  column\/}). 
  The cuts have been taken along the long axis of ROI 1, averaging over
  the central 5\,pixels across that axis.
  {\it Top row\/}: non-LTE-inferred field.
  {\it Bottom row\/}: MHS field.
  The non-LTE and MHS panels have been clipped to the same values according the
  colour bars at the top of the figure.
  The dashed lines indicate the heights corresponding to $\ltau=-5.1$ ({\it
  upper line\/}) and $-$4.1 ({\it lower line\/}) in the upper panels and the
  equivalent column mass in the MHS model in the lower panels, where the latter
  panels also show the field over a larger $z$-height range than in the top row.
  }
\end{figure*}

\subsection{Detailed comparison of selected regions-of-interest}
\label{sec:rois}
\subsubsection{ROI 1: Sunspot vicinity}
\label{sec:roi1}
Figure~\ref{fig:roi1} presents a comparison of the observations, inversions and
modelling results for the detailed view of ROI 1, just north of the negative
polarity sunspot.
The first eight columns compare the intensity maps in all four Stokes
components from the observations and best fits as determined by the non-LTE
inversions and serve primarily to assess the quality of the inferred model
atmosphere.
The remaining nine columns show the \Blos, \Bhor\ and \Bazi\ from the MHS model,
WFA and non-LTE STiC inversions. 
The contours in all panels highlight the regions where the WFA-derived \Bhor\
reaches 150\,G (solid contours) and 350\,G (dashed). 
As the ROI has been rotated counter-clockwise by 90\deg, the Solar-X axis lies
along the vertical axis and the sunspot lies to the right of these cut-outs, up
to about $X=74\arcsec$.

The magnetic field maps have been averaged over $\ltau=-4.1$ and $-$5.1, since
inspection of the response functions of \CaII\ Stokes $Q$, $U$ and $V$ inidcates
maximum sensitivity to the magnetic field in that \ltau-depth range.
The MHS maps are averaged over the equivalent range in column mass, as derived
from the non-LTE inversions (see Section~\ref{sec:mhs_postproc}).

Comparing the observed and synthetic maps, we see that especially Stokes $I$ and
in general also $V$ are well-reproduced.
The synthetic map for the latter contains locations where strong signal is more
extended compared to the observations (\eg\ the blue patch around
$X=66\arcsec$), but also places where signal is lacking (\eg\ around
$X=72\arcsec$).
This translates into \Blos\ maps that in their general distribution of field
concentrations are essentially the same between the MHS model, WFA results and
the non-LTE inversions.
However, both the WFA and especially the non-LTE inversions return stronger
field in the chromosphere, \eg\ close to the sunspot (between $X=66$\arcsec and
73\arcsec) and between $X=80$\arcsec and 82\arcsec. 

The non-LTE inversions struggle somewhat more for Stokes $Q$ and $U$, though
mostly where the horizontal field is expected to be weak (based on the WFA).
For instance, within the contour lines centred at $X=77\farcs{5}$, where the
WFA-derived \Bhor\ is less than 150\,G, and and in general above $X=80\farcs{5}$
the synthetic maps hardly recover any signal even though the observed maps
clearly exhibit some structure.
This is reflected in both the inferred \Bhor\ and \Bazi\ maps, that are
particularly noisy between $X=75$\arcsec\ and 83\arcsec.

Figure~\ref{fig:vcut_roi1} shows vertical cuts in \Blos\ and \Bhor\ along
the long axis of ROI 1, averaged over the central 5 pixels across that axis, for
both the non-LTE-inferred field and the MHS field.
The dashed lines indicate the equivalent $\ltau=-4.1$ and $-$5.1 depths over
which the field has been averaged for display in Fig.~\ref{fig:roi1}, where the
$\ltau=-5.1$ depth is more corrugated than $\ltau=-4.1$, especially in the
inversions (upper row).
Both the \Blos\ and \Bhor\ cuts look very similar between the non-LTE
inversions and the MHS model.
There are \Blos\ field concentrations rooted at $z=0$\,Mm around $X=68$\arcsec,
70\farcs{5} and 82\arcsec, although the non-LTE inversions have the strong field
extending further and persist between about 0.6--1.2\,Mm, while the MHS field
exhibits a much more diffuse distribution at those heights.
Similarly, the \Bhor\ cuts have a pair of stronger field concentrations around
$X=80\farcs{5}$ and 82\arcsec\ that in both cases extend up to about
$z=0.6$\,Mm in the non-LTE panel and slightly lower in the MHS panel.
The marked difference is stronger transverse field above 0.5\,Mm between
$X=65$\arcsec\ and 67\arcsec\ for the non-LTE inversions, that is absent in the
MHS model.
Finally, the lower atmosphere (up to about 0.5--0.6\,Mm in the non-LTE case and
up to about 1\,Mm in the MHS case) between $X=74$\arcsec\ and 79\arcsec\
exhibits close-to-zero \Blos\ and \Bhor\ field strengths in both the inversions
and MHS model.

\begin{figure*}[bht]
  \centerline{\includegraphics[width=\textwidth]{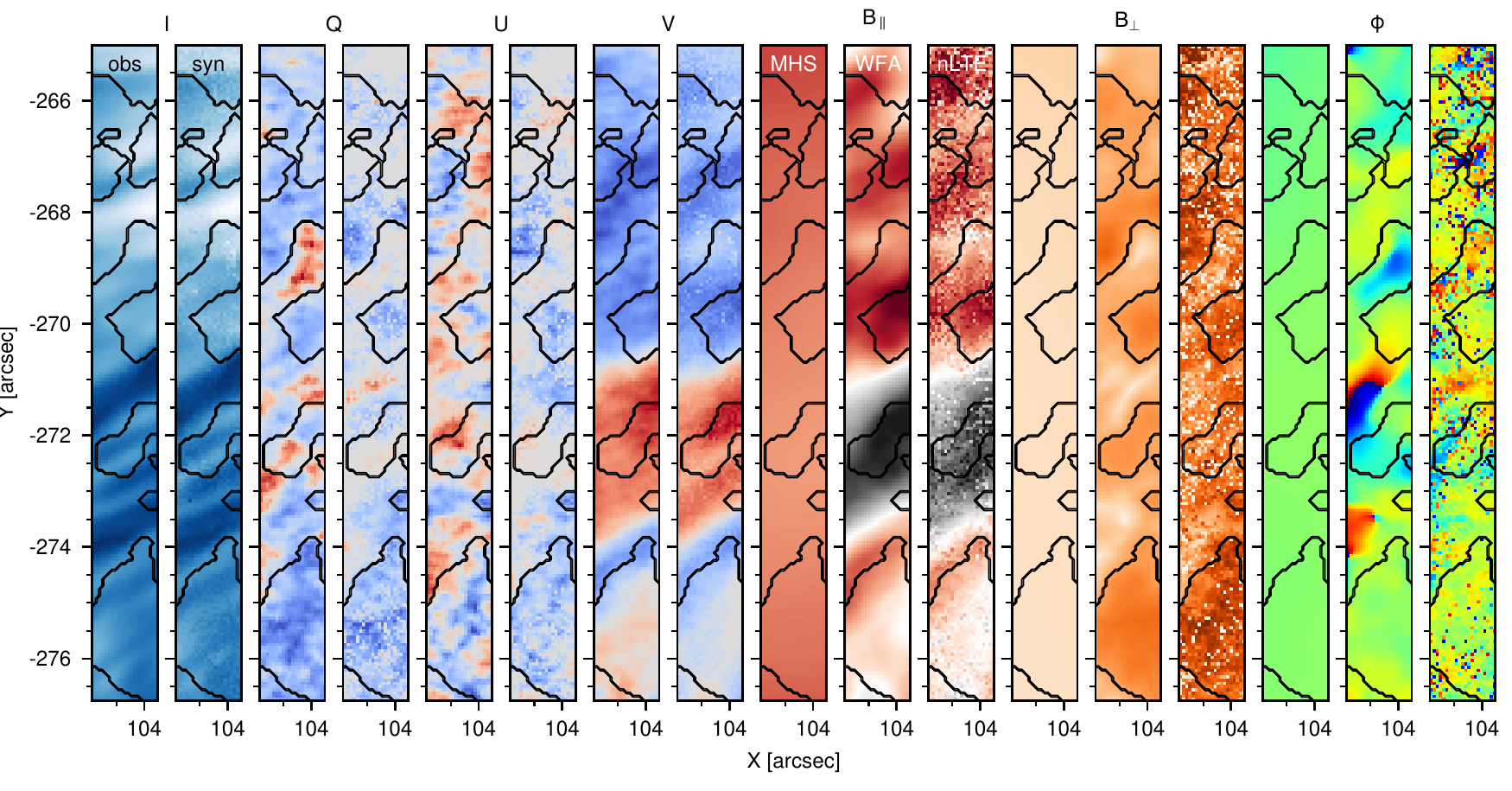}}
  \vspace{-2.5ex}
  \caption[]{\label{fig:roi3} %
  Maps of Stokes intensity and magnetic field components for ROI 3.
  Format as for Fig.~\ref{fig:roi1}, except that contours are only
  shown for $\Bhor=150$\,G in the WFA-derived field.
  }
\end{figure*}

\begin{figure*}[bht]
  \centerline{\includegraphics[width=\textwidth]{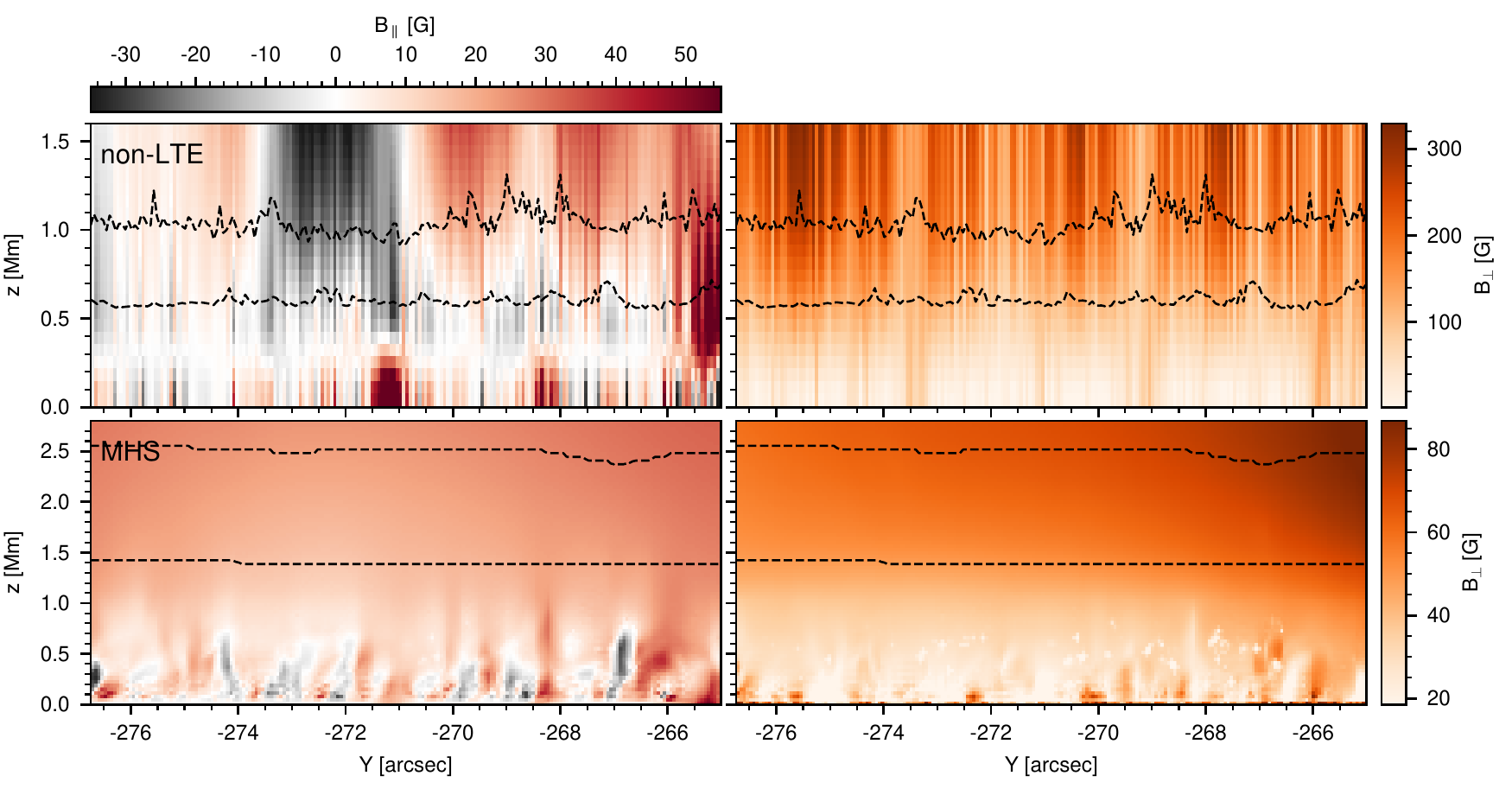}}
  \vspace{-2.5ex}
  \caption[]{\label{fig:vcut_roi3} %
  Vertical cuts in \Blos\ ({\it left column\/}) and \Bhor\ ({\it right
  column\/}) for ROI 3. 
  Format as for Fig.~\ref{fig:vcut_roi1}, except that the \Bhor\ panels ({\it
  right column}) have been clipped individually, according to the colour bars on
  the right of the figure.
  }
\end{figure*}

\subsubsection{ROI 3: High fibrils}
\label{sec:roi3}
Figures~\ref{fig:roi3} and \ref{fig:vcut_roi3} present the results for ROI 3
in the same format as Figs.~\ref{fig:roi1}--\ref{fig:vcut_roi1} for ROI 1.
The location for ROI 3 was chosen for its crossing of a set of the longest
fibrils in our mosaic observations, that are evident as dark slanted bands
around $Y=-272$\arcsec\ in Stokes $I$ (Fig.~\ref{fig:roi3}, first two columns).
Stokes $V$ is particularly strong at the location of these fibrils, which is
reflected in a seemingly homogeneous band of strong negative polarity in both
WFA- and non-LTE-inferred \Blos\ maps.
While both Stokes $Q$ and $U$ show structure in their signature at this
location, the correlation with the dark bands in Stokes $I$ is not evident.
Still, the WFA recovers \Bhor\ in excess of 150\,G in a patch overlapping with
the strongest negative \Blos\ at the location of these fibrils (cf.~contours at
$Y=-272$\arcsec).

Unsurprisingly, the non-LTE inversions struggle in reproducing the $Q$ and $U$
profiles in the areas where the WFA already suggests the transverse field is
weaker (cf.~the $\Bhor=150$\,G contours, for instance between $Y=-274\farcs{5}$
and $-$272\farcs{5} and between $Y=-272$\arcsec\ and $Y=-269$\arcsec), while the
stronger field contour-enclosed areas typically show a similar amount of
structure between the observed and synthetic $Q$ and $U$ maps.
Notwithstanding, the non-LTE \Bhor\ agrees qualitatively with the WFA-inferred
\Bhor\ for most of the sub-FOV, albeit with more noise.
The same holds for the azimuth \Bazi, with the exception of several smaller
patches that are found in either the WFA or non-LTE inversions (\eg\ the
blue-red patch at $Y=-271\farcs{5}$ in the WFA azimuth that is not at all
well-defined in the non-LTE results).

One stark difference when comparing the field from different methods, is that
the MHS results show little to no field structuring at the \ltau-equivalent
heights used to construct these maps, in contrast to ROI 1 (and to some extent
also ROI 2, see Appendix~\ref{sec:roi2}).
In particular, the strong \Blos\ (and to certain extent \Bhor) associated with
the dark fibrils in the \CaII\ line core intensity image is absent in the MHS
model and instead the MHS field components exhibit little to no structure.
These discrepancies are also reflected in the vertical cross-cuts shown in
Fig.~\ref{fig:vcut_roi3}, where at the height of the horizontal cuts (\ie\
between the dashed lines) the MHS model exhibits a weakly positive \Blos\ of
less than 30\,G and \Bhor\ below 80\,G in the strongest-field part around
$Y=-266$\arcsec.
While at the equivalent \ltau-depths the non-LTE-inferred \Blos\ is only barely
stronger in absolute terms compared to the MHS results, \Bhor\ is stronger by up
to a factor 5 in the top right non-LTE panel.

Lower down in the atmosphere, below 500\,km, the alternating positive and
negative \Blos\ polarities in the MHS panel, and in particular their extensions
to 700--800\,km, are not always as well-recovered in the non-LTE results.
However, careful inspection shows that most of the stronger-field concentrations
at heights 0--200\,km are located at approximately the same place in both
panels, albeit typically with larger field strengths in the non-LTE case, \eg\
the positive polarities at $Y=-273\farcs{8}$, $-$272\farcs{5}, between
$-$271\farcs{5} and $-$271\arcsec, and between $-$268\farcs{5} and
$-$268\arcsec, or the negative polarities at $Y=-276$\arcsec, near
$-$274\arcsec\ and $-$267\farcs{7}.

A similar structuring of alternating stronger and weaker \Bhor\ along the
vertical cut is found in the lower atmosphere of the MHS model, while the
non-LTE cut only shows the stronger concentrations of those, \eg\ at
$-$273\farcs{5}, $-$269\arcsec\ and $-$266\arcsec.
It turns out the inversions of \FeI\ struggled, especially for Stokes $Q$ and
$U$.
Higher up in the atmosphere, above about 1\,Mm, the results diverge even
further, with a \Bhor\ concentration in excess of 300\,G in the non-LTE panel
around $Y=-276$\arcsec, while the MHS \Bhor\ is a factor 5--6 weaker at
equivalent column mass.
On the other hand, the \Bhor\ values in the upper right corner differ only by up
to a factor 2 and are often even quite close between non-LTE and MHS results.

\section{Discussion}
\label{sec:discussion}
\subsection{Chromospheric magnetic field on active region scales}
\label{sec:discussion_fov}

Qualitatively, the MHS and WFA field agree in spatial distribution of the
stronger-field concentrations, \ie\ sunspots and pores, but the WFA shows more
enhanced field fine-structure (cf.~Fig.~\ref{fig:wfa_mhs}), which can be most clearly
seen for the transverse component in the superpenumbra of either sunspot.
While a contribution from the photosphere could conceivable play a role, it is
unlikely the case here as both Stokes $Q$ and $U$ exhibit this strong-signal
superpenumbral fine-structure already in the first off-core wavelength samplings
and the restricted wavelength range chosen to determine the WFA with would
furthermore minimise any photospheric contribution.

Quantitatively, \Blos\ exhibits a tighter correlated between the WFA and MHS
model than \Bhor, while the latter is closer to a one-to-one relationship, yet
in both cases the WFA field strengths tend to be larger than in the MHS model.
This tendency can easily be understood by considering the maps in
Fig.~\ref{fig:wfa_mhs}, which show a much more extended region of strong-field
\Blos\ for the sunspots in the WFA map than in the MHS map, especially for the
positive-polarity sunspot.
Similarly, the strongest enhancements in \Bhor\ are found for the superpenumbral
substructure in the WFA map, while the MHS exhibits a much smoother spatial
distribution, except for enhanced field crossing the sunspot centres (which
is data processing artefacts).
The latter have no counterpart in the WFA map, explaining the scatter cloud in the
top left of the right-hand panel in Fig.~\ref{fig:wfa_mhs_scatter}, while the
former are the reason for the majority of the distribution lying to the right of
the one-to-one diagonal.

On large scales the MHS azimuth appears to follow the \CaIR\ fibril
structures more closely than the WFA azimuth does and this is particularly so
outside the sunspots and pores. 
There the \CaII\ Stokes $Q$ and $U$ signals are weak and noisy, leading to
weaker and more uncertain inferred transverse field strengths and azimuths, and
the visible misalignment between fibrils and WFA azimuth could therefore, at
least in part, be attributable to the signal-to-noise ratio of the underlying
observations.
However, whether the inferred field should follow the line-core fibril
structures may depend on the target.
For instance, 
\citetads{2011A&A...527L...8D} 
found most of the derived azimuths to be consistent with the fibril orientation
in two targets observed with \CaIR\ spectropolarimetry. 
Similarly, 
\citetads{2013ApJ...768..111S}, 
found the
inferred field orientation to be well-aligned with the projected angles of the
fibrils, to within an error of only 10\deg, and
\citetads{2017A&A...599A.133A} 
report alignment to within about 16\deg\ for the superpenumbra, although
the misalignment could be up to about 34\deg\ in magnetically weaker plage regions.
We observe a qualitatively similar effect, in that the WFA azimuth appears to
trace the dark superpenumbral fibrils reasonably well, but is more often at odds
with fibrils outside the stronger-field concentrations.

\subsection{Magnetic field from non-LTE inversions}
\label{sec:discussion_rois}
In general, the non-LTE inversions recover a similar spatial distribution of the
chromospheric field as the WFA, in particular for the \Blos\ component, but to a
decent extent also for \Bhor (at least where the transverse field is relatively
strong).
Unsurprisingly though, the inversions struggle to derive clean maps for \Bhor\
wherever Stokes $Q$ and/or $U$ are relatively noisy (\eg\ ROIs 3 and 4 in
Figs.~\ref{fig:roi3} and \ref{fig:roi4}, respectively). 
Despite the spatially-regularised WFA having served as initial guess, the \Bhor\
maps for ROIs 3 and 4 (and in part also ROI 2) are far from smooth and this
carries over into the non-LTE-inferred azimuth that for most of those FOVs
disagrees with the WFA-derived azimuth even in a broad sense
(cf.~Figs.~\ref{fig:roi2} and \ref{fig:roi4} for ROIs 2 and 4, respectively). One reason for the increased noise is that in order for non-LTE inversions to work, the stratification of all parameters must be inferred correctly, not only the magnetic field. This leads to more degenerate solutions and hence more noise in the inferred parameters.

Of all ROIs, the first shows the highest degree of similarity between the MHS,
WFA and non-LTE inversions for all three magnetic field components.
ROI 2 comes a close second, at least for \Blos\ and \Bhor, while ROIs 3 and 4
differ considerably in most aspects.
Indeed, while all four ROIs exhibit similar photospheric patterns between the
MHS model and non-LTE inversions in the vertical cross-cuts for \Blos\ and
\Bhor, the MHS \Bhor\ field strengths in ROIs 3 and 4 are up to a factor 5
smaller at chromospheric heights than what the non-LTE inversions indicate.
This can be understood from the proximity to strong-field concentrations for
ROIs 1 and 2 persisting from the photosphere to greater heights
(cf.~the vertical cross-cuts in Figs.~\ref{fig:vcut_roi1} and
\ref{fig:vcut_roi2}), in contrast to ROIs 3 and 4 that are located in more
`quiet' parts of the FOV.
Given that the MHS model extrapolates from the photospheric field, it will
inherently exhibit a better agreement at chromospheric heights in the vicinity
of strong-field concentrations than in magnetically weaker regions.

Also, despite the large difference in transverse chromospheric field strengths
in the `quiet' ROIs, the distribution of weaker field in the photosphere and
stronger field in the chromosphere is qualitatively similar between the MHS
model and non-LTE inversions.
This pattern can be explained by the difference in origin of the field.  
In the lower atmosphere the field comes from the (weak) photospheric polarities,
while the field higher up is that connecting the active region's main
polarities.
As such, both the MHS model and non-LTE inversions present a similar magnetic
field stratification of the active region's connectivity, only that the latter
recovers stronger transverse field in the chromosphere.

Evidently, ROI 3 presents a further discrepancy that cannot be explained in a
similar way.
Here, both the WFA and non-LTE inversions pick up on the enhanced \CaIR\ Stokes
polarisation corresponding to the dark fibrils traversing this sub-field, while
the MHS model shows essentially a smooth extension from the upper-photospheric
field state.
Despite that these dark fibrils appear to connect both sunspots and are thus
expectedly rooted in either of the main polarities in the photosphere, the MHS
model is unable to reproduce the magnetic imprint thereof at chromospheric
heights.

\section{Conclusions}
\label{sec:conclusions} 
We have presented an analysis of the chromospheric magnetic field in AR 12723 as
inferred with a spatially-regularised weak-field approximation and a non-LTE
inversion code, and compared these against a magnetohydrostatic model based
solely on the photospheric vector magnetic field.
We find that the chromospheric field is qualitatively similar between these
methods, where the line-of-sight component is best reproduced in the MHS model,
while the transverse component is consistently weaker than the WFA-derived
field, especially outside the sunspots. 
In general, the MHS model also presents a smoother field component maps than the
WFA, despite the spatial regularisation applied in the latter, and while the MHS
model is unable to recover the magnetic imprint in \Blos\ and \Bhor\ from a set
of long fibrils that appear to connect both sunspots, there is otherwise
remarkably good agreement with the WFA.

In addition, the vertical cuts in magnetic field offer a similar view in the
MHS model and non-LTE inversions, especially for the two ROIs that are close to
the stronger field concentrations (\ie\ the negative-polarity sunspot and
crossing the pores) and at least qualitatively for the ROIs that are taken in
more ``quiet'' parts of the field-of-view, albeit in some places with transverse field
strengths up to a factor 5 or so weaker than in the non-LTE inversions or WFA.
On the other hand, the non-LTE inversions struggle in recovering the transverse
component when its strength is below some 200--300\,G.
For the more quiet ROIs the non-LTE inversions also appear to ``overshoot'' and
settle on stronger \Bhor\ than the WFA indicates, even though the Stokes $Q$ and
$U$ are not always recovered in magnitude.
The noisiness of the non-LTE-inferred \Bhor\ and \Bazi\ maps in contrast to
those from the spatially-regularised WFA further emphasis the need for spatial
coupling in the inversions as proposed in recent years by 
\citeads{2015A&A...577A.140A}, 
\citeads{2019A&A...631A.153D}). 

While the MHS model agrees qualitatively (and to a large extent quantitively)
with the observationally inferred chromospheric field, it does not retrieve all
the features from the observations and our results therefore support earlier
studies which indicate that including a
chromospheric constraint in field extrapolations or data-driven modelling can
improve the recovery of chromospheric and coronal magnetic structures.
%
In particular the difficulty in reaching a transverse chromospheric field
strength similar to that observationally inferred or recovering the presence of
high fibrils suggests that key ingredients may thus still be missed, whereas an
accurate estimate of the chromospheric
field within, for instance, erupting flux ropes is vital to properly predict the
geo-effectiveness of earth-bound CMEs 
\citepads{2019SpWea..17..498K}. 

\begin{acknowledgements}
%
GV has been supported by a grant from the Swedish Civil Contingencies Agency (MSB).
This research has received funding from the European Union's Horizon 2020
research and innovation programme under grant agreement No 824135.
JL and SD are supported by a grant from the Knut and Alice Wallenberg foundation
  (2016.0019).
  XZ is supported by the mobility program (M-0068) of the Sino-German Science Center.
This project has received funding from the European Research
  Council (ERC) under the European Union's Horizon 2020 research and innovation
  programme (SUNMAG, grant agreement 759548).
This project has received funding from the European Research Council (ERC) under
  the European Union's Horizon 2020 research and innovation programme (SolMAG,
  grant agreement No 724391). 
The Institute for Solar Physics is supported by a grant for research
infrastructures of national importance from the Swedish Research Council
(registration number 2017-00625).
The Swedish 1-m Solar Telescope is operated on the island of La Palma by the
Institute for Solar Physics of Stockholm University in the Spanish Observatorio
del Roque de los Muchachos of the Instituto de Astrof{\'i}sica de Canarias. 
The SDO/HMI data used are courtesy of NASA/SDO and HMI science team.
The inversions were performed on resources provided by the Swedish National
Infrastructure for Computing (SNIC) at the National Supercomputer Centre at
Link{\"o}ping University.
We made much use of NASA's Astrophysics Data System Bibliographic Services.
Last but not least, we acknowledge the community effort to develop open-source
  packages used in this work: 
  {\tt{numpy}} (\citeads{oliphant2006guide}; \url{numpy.org}), 
  {\tt{matplotlib}} (\citeads{Hunter:2007}; \url{matplotlib.org}), 
  {\tt{scipy}} (\citeads{2019arXiv190710121V}; \url{scipy.org}),
  {\tt{astropy}} (\citeads{astropy:2013}, \citeads{astropy:2018}; \url{astropy.org}),
  {\tt{sunpy}} (\citeads{sunpycommunity2020}; \url{sunpy.org}),
  {\tt{scikit-learn}} (\citeads{scikit-learn}; \url{scikit-learn.org}).

\end{acknowledgements}

\bibliographystyle{aa}

\bibliography{msb_mosaic} 

\begin{appendix}
\section{Additional regions of interest}
\label{sec:other_rois}
In this Appendix we present the results for additional regions of interest, in
similar format as for ROIs 1 and 3.

\begin{figure*}[bht]
  \centerline{\includegraphics[width=\textwidth]{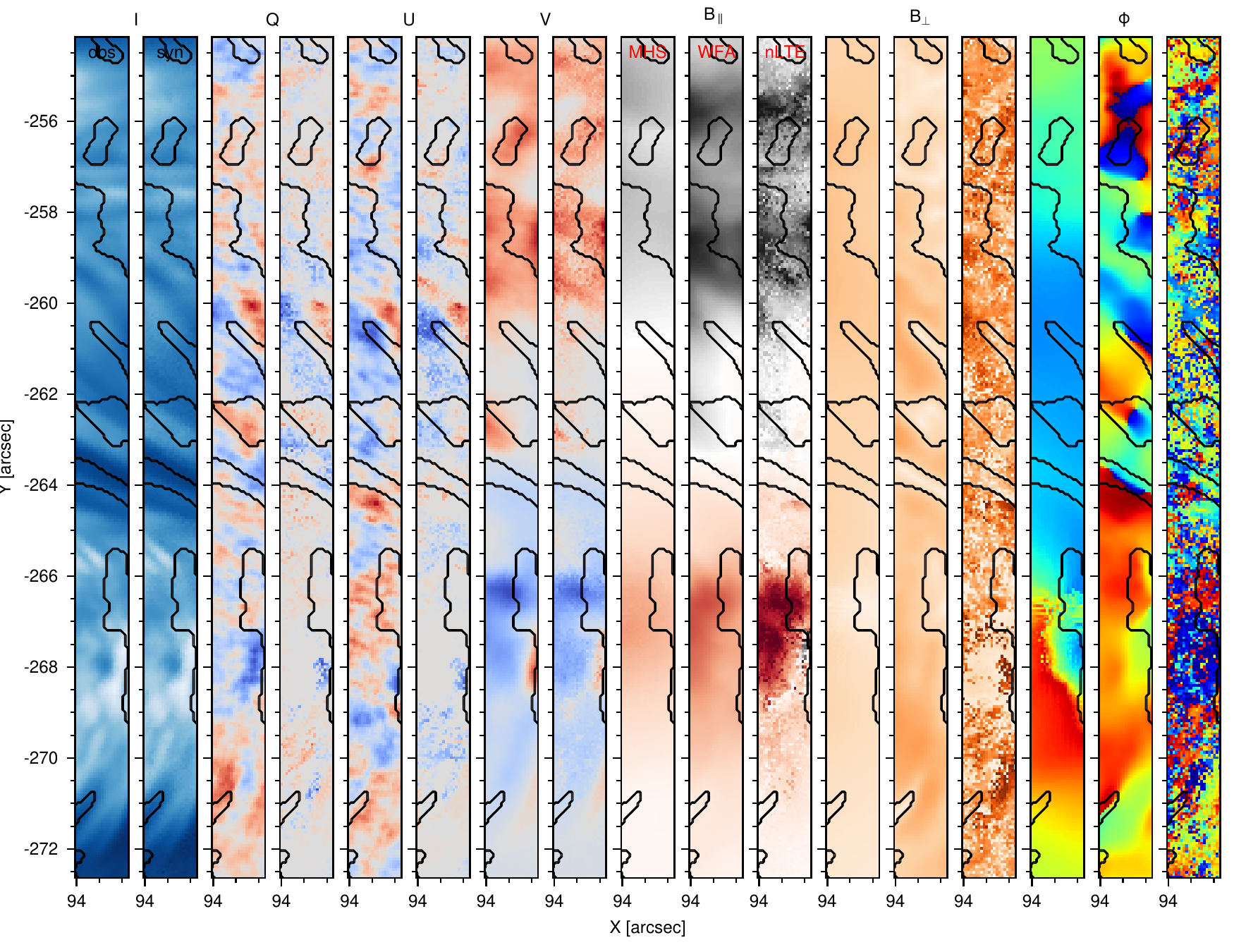}}
  \vspace{-2.5ex}
  \caption[]{\label{fig:roi2} %
  Maps of Stokes intensity and magnetic field components for ROI 2.
  Format as for Fig.~\ref{fig:roi1}, except that the contours now only show the
  WFA-derived $\Bhor=150$\,G boundary.
  }
\end{figure*}

\begin{figure*}[bht]
  \centerline{\includegraphics[width=\textwidth]{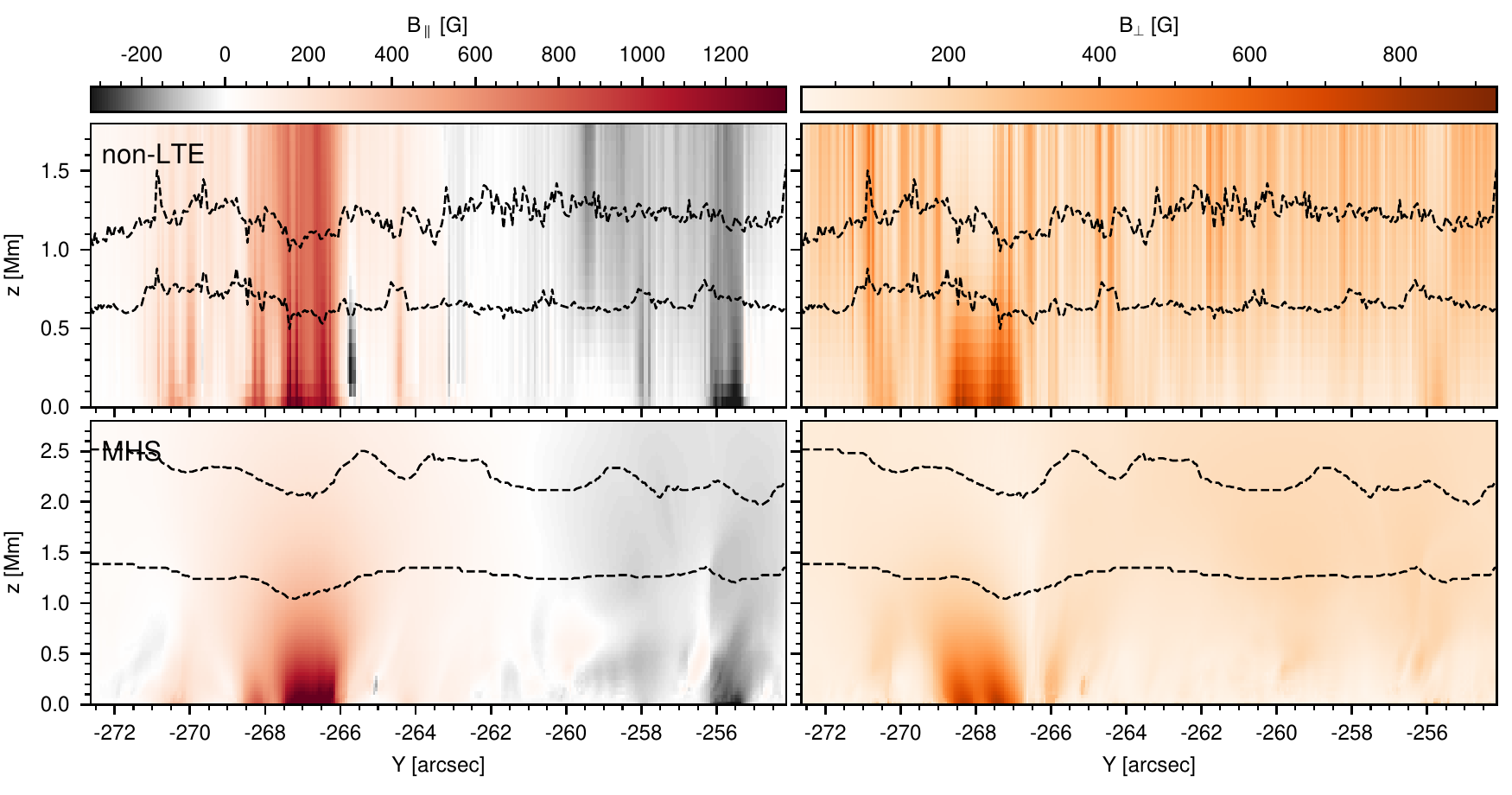}}
  \vspace{-2.5ex}
  \caption[]{\label{fig:vcut_roi2} %
  Vertical cuts in \Blos\ ({\it left column\/}) and \Bhor\ ({\it right
  column\/}) for ROI 2, in the same format as Fig.~\ref{fig:vcut_roi1}.
  }
\end{figure*}

\subsection{ROI 2: Pore and strong-field concentrations}
\label{sec:roi2}
The figure pair Fig.~\ref{fig:roi2} and \ref{fig:vcut_roi2} presents the
results for ROI 2 in a similar format as Figs.~\ref{fig:roi1} and
\ref{fig:vcut_roi1}, only with the contours in Fig.~\ref{fig:roi2} now
indicating solely the $\Bhor=150$\,G level.
Similar to ROI 1, Stokes $I$ and $V$ are in general best-reproduced, while $Q$
and $U$ are more problematic, especially where the WFA already indicates that
the field is weak.
For instance, the regions to the right of the contours around \XYis{95}{-266},
the band around $Y=-264$\arcsec, or within the contours around
$Y=-262\farcs{5}$ are mostly devoid of signal in the
synthetic $Q$ and $U$ maps.
Surprisingly, this is also the case for that part of the ROI below the contour
around $Y=-264$\arcsec, where the WFA returned stronger \Bhor\ and one could
have expected a similar result as above the contour at $Y=-262$\arcsec\ and to
the left of those crossing the vertical edges at $Y=-259\farcs{5}$\arcsec\ and
$-$257\farcs{5}, where much of the structure in linear polarisation is
reproduced.

This ROI crosses a positive polarity pore around $Y=-267$\arcsec\ and
another stronger field concentration around $Y=-256$\arcsec, as is evident
from the \Blos\ maps in both Figs.~\ref{fig:roi2} and \ref{fig:vcut_roi2}.
Especially around the pore, both MHS, WFA and non-LTE inversions agree in the
top-down viewed shape of the positive polarity, although the non-LTE inversions
suggest much stronger field at those heights than either the WFA and especially
the MHS results.
Also, where the WFA and non-LTE inversions return strong negative \Blos\ (\ie\
around $Y=-258$\arcsec\ and especially $-$256\arcsec), there no such significant
concentration in the MHS model.
Looking for the same locations in the vertical cut, we find that the MHS model
does have marked \Blos\ field concentrations at those $Y$ locations, yet already
strongly atenuated at heights equivalent to $\ltau=-4.1$ (lower dashed line) and
field strengths of some 500--600\,G above the pore and $-$100\,G around
$Y=-256$\arcsec.
In contrast, the non-LTE inversions have field strengths of, respectively, order
800\,G and $-$200\,G persisting above $\ltau=-4.1$.
In both cases, \Blos\ is strongly vertical above the positive polarity pore, but
somewhat more inclined with height in the MHS model for the negative polarity.

The transverse field shows the strongest concentrations at the location of the
pore, only extending a few hundred kms in height in the MHS model and reduced to some
200--300\,G between the $\ltau=-4.1$ and $-$5.1 lines, while the non-LTE
  inversions exhibit \Blos\ in excess of 400\,G around $Y=-267\farcs{5}$.
For most of the vertical cut, both non-LTE inversions and MHS model exhibit weak
\Bhor\ field.

Where the azimuth for ROI 1 is largely similar between MHS, WFA and non-LTE
inversions for most of that sub-FOV, this is not the case for ROI 2 (last three
columns of Fig.~\ref{fig:roi2}).
Indeed, although we find some resemblance between MHS and WFA azimuths south of
about $Y=-268$\arcsec, only few other spots in the FOV are anywhere near
agreement (\eg\ close to $Y=-260\farcs{5}$, $-$258\arcsec\ or $-$254\farcs{5}).
Noise dominates the non-LTE azimuth, althoguh with some effort similar patterns
as in the WFA azimuth can be found south of $Y=-269$\arcsec, between the
contours at $Y=-263\farcs{5}$ and $-$263\arcsec, between $Y=260\farcs{5}$ and
$-$258\farcs{5} and north of $Y=-255\farcs{5}$.

\subsection{ROI 4: Fibrils in sunspot vicinity}
\label{sec:roi4}
\begin{figure*}[bht]
  \centerline{\includegraphics[width=\textwidth]{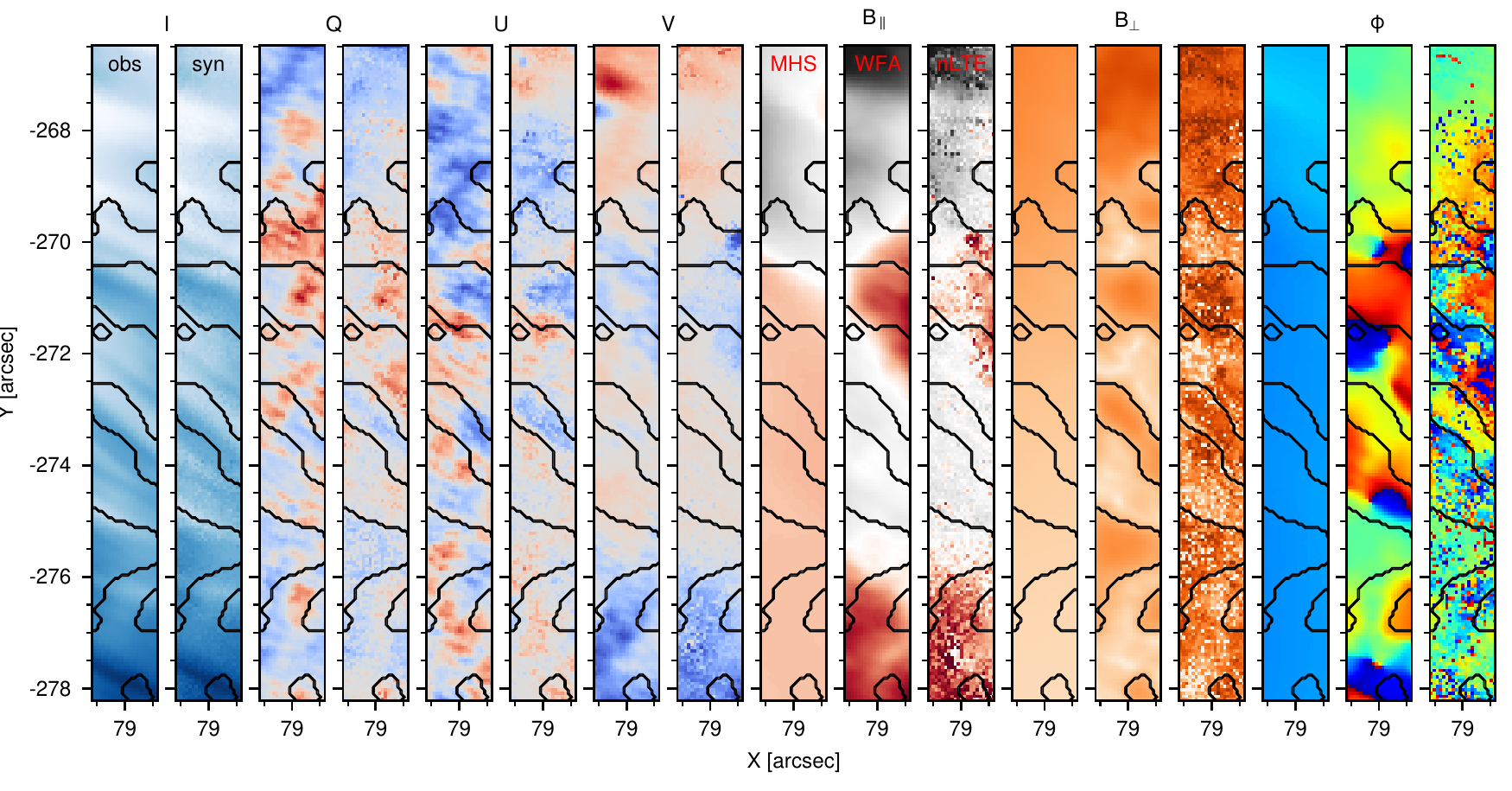}}
  \vspace{-2.5ex}
  \caption[]{\label{fig:roi4} %
  Maps of Stokes intensity and magnetic field components for ROI 4.
  Format as for Fig.~\ref{fig:roi2}.
  }
\end{figure*}

\begin{figure*}[bht]
  \centerline{\includegraphics[width=\textwidth]{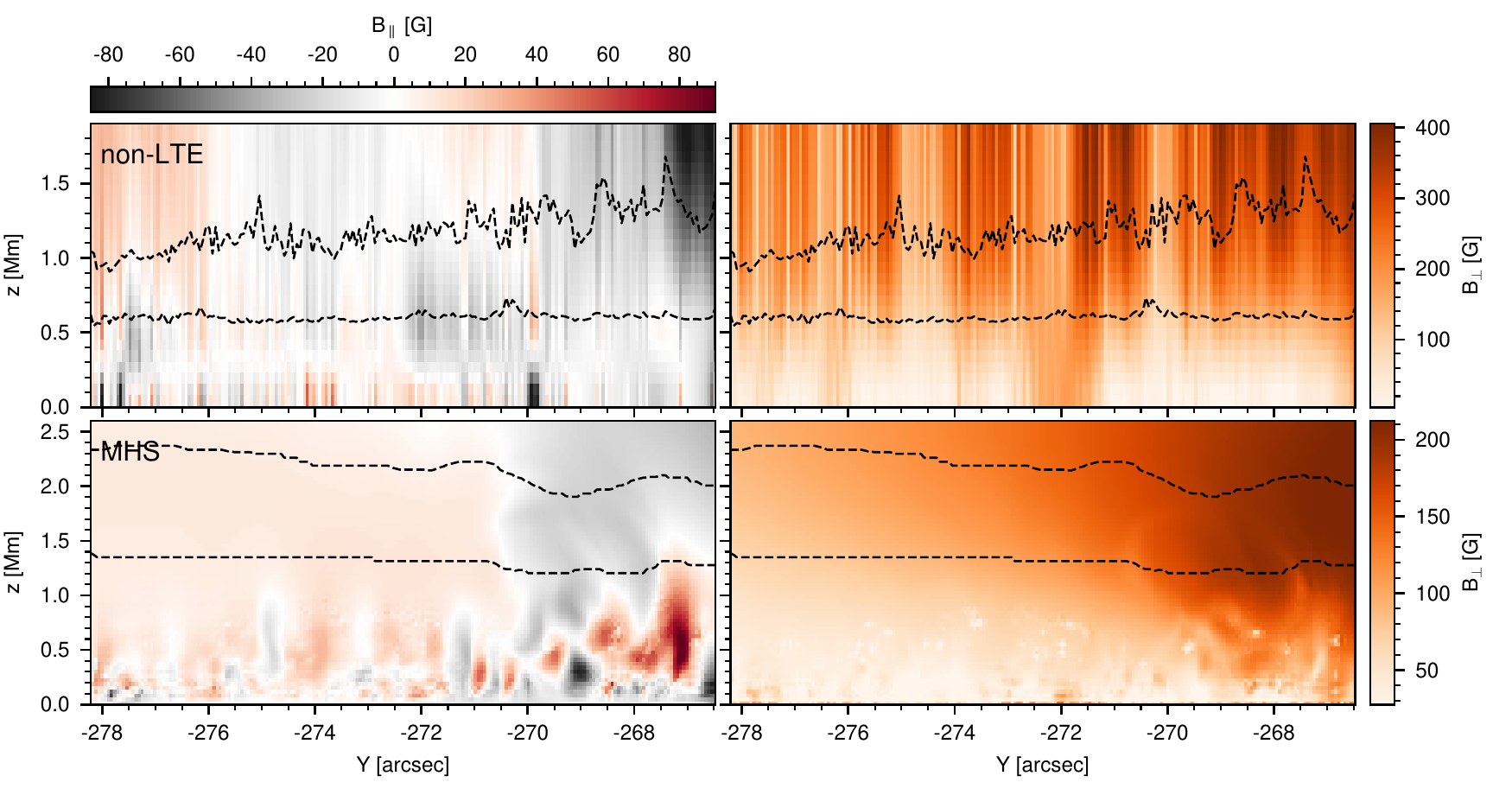}}
  \vspace{-2.5ex}
  \caption[]{\label{fig:vcut_roi4} %
  Vertical cuts in \Blos\ ({\it left column\/}) and \Bhor\ ({\it right
  column\/}) for ROI 4, in the same format as
  Fig.~\ref{fig:vcut_roi3}.
  }
\end{figure*}

Finally, Figs.~\ref{fig:roi4} and \ref{fig:vcut_roi4} present the results for
a region covering some of the longer fibrils extending to the south-west of the
negative-polarity sunspot.
Similar to ROI 3, Stokes $Q$ and $U$ are again best reproduced where the WFA
indicates $\Bhor \geq 150$\,G (dark orange within contours in the middle \Bhor\ panel),
although the magnitude of the signal is not always reproduced (\eg\ red in
Stokes $U$ at $Y=-275\farcs{5}$ or blue in the same panel but around
$-$267\farcs{5}).
Outside these contours, little signal is found in the synthetic $Q$ and $U$
profiles and this explains the noisiness of the \Bhor\ and \Bazi\ at those
locations.
Within the stronger-\Bhor\ contours the non-LTE-inferred \Bhor\ and \Bazi\
appear more homogeneous, with the former reaching larger field strengths than in
the WFA, while the corresponding \Bazi\ panels show largely the same angle
in those places, except within the contour at $Y=-271$\arcsec.
Interestingly, the non-LTE azimuth much more clearly resembles the WFA results
compared to ROI 2, even though the \Bhor\ magnitudes between the two ROIs are
similar.

The MHS model exhibits much weaker field throughout, of the order of the weaker
field in the WFA cut, although qualitatively it agrees in having stronger \Bhor\
in the upper part of the ROI, north of about $Y=-270$\arcsec.
This is also reflected in the vertical cross-cut (Fig.~\ref{fig:vcut_roi4},
right column), showing an area of stronger field in the upper right corner of
the cross-cut for both non-LTE inversions and MHS model.
The \Bhor\ in the lower atmosphere shows little resemblance between the two,
however, except maybe the location of a region of stronger \Bhor\ around
$Y=-272$\arcsec, albeit with a much smaller extent in the MHS model.
Finally, the MHS azimuth \Bazi\ (Fig.~\ref{fig:roi4}, third column from the
right) is nowhere near close that of the WFA or non-LTE inversions.

Meanwhile, Stokes $V$ is largely reproduced, except some of the details around
$Y=-270$\arcsec\ and $-$267\arcsec, yet this seems to have little effect on the
determined \Blos\ when compared to the WFA.
The contrast with the MHS \Blos\ is evident, however.
The band of near-zero negative polarity recovered in the WFA and non-LTE
inversions between $Y=-275\farcs{5}$ and $-$271\farcs{5} (correlating with a
likewise weakly positive Stokes $V$ signal) is positive in the MHS model and the
stronger negative polarity in the former two around $-$267\arcsec\ is barely
negative in the MHS model.
The vertical cross-cut (Fig.~\ref{fig:vcut_roi4}, left column) shows the latter
is because of a transition from positive to negative polarity where a strong
positive field concentration extending over roughly 1\,Mm and into the bottom
part of the height range used for the maps in Fig.~\ref{fig:roi4}.
Also, any sign of a band of weak negative \Blos\ left of
$Y=-271\farcs{5}$ is missing in the MHS model.
On the other hand, the \Blos\ in the lower atmosphere of the non-LTE inversions
bears some similarity to that in the MHS results, at least left of about
$Y=-270$\arcsec.
Even though generally not agreeing in magnitude and often not in (vertical)
extent either, the negative polarity concentrations close to $Y=-278$\arcsec,
$-$277\farcs{5}, and near $-$270\arcsec, as well as the positive concentrations
between $Y=-274\farcs{5}$ and $-$271\farcs{5} or the alternating
positive-negative concentrations between $Y=-271\farcs{5}$ and $-$270\arcsec\
are similar between the two panels.

\end{appendix}

\end{document}